\documentclass[10pt,letterpaper]{article}
\usepackage[top=2.5cm,bottom=2.5cm,left=2.5cm,right=2.5cm]{geometry}
\usepackage{graphicx}
\usepackage{float}
\usepackage{subcaption}
\usepackage{multirow, multicol, makecell, booktabs}
\usepackage{hhline}
\usepackage{braket}
\usepackage{authblk}
\usepackage{soul}
\usepackage{algorithm}
\usepackage{algpseudocode}
\usepackage{tikz}
\usepackage{xcolor}
\usepackage{amsmath}
\usepackage{amssymb}

\usepackage{hyperref}
\hypersetup{
     colorlinks = true,
     citecolor  = blue,
     linkcolor = blue,
     urlcolor = blue
}

\usepackage[authoryear,round]{natbib}
 \bibpunct[, ]{(}{)}{,}{a}{}{,}%
\usepackage{url}

\DeclareRobustCommand{\VAN}[3]{#2}  

\newcommand{\Reach}{{\sc Reach}}
\newcommand{\MartaReach}{{\sc MARTA Reach}}

\title{MARTA Reach: Toward an On-Demand Multimodal Transit System in Atlanta}

\author[1]{Pascal~Van~Hentenryck \footnote{Email: \href{mailto:pvh@gatech.edu}{pvh@gatech.edu}}}
\author[1]{Connor~Riley}
\author[1]{Anthony~Trasatti}
\author[1]{Hongzhao~Guan}
\author[1]{Tejas~Santanam}
\author[1]{Jorge~A.~Huertas}
\author[1]{Kevin~Dalmeijer}
\author[2]{\\ Kari~Watkins}
\author[3]{Juwon~Drake}
\author[4]{Samson~Baskin}

\affil[1]{H. Milton Stewart School of Industrial and Systems Engineering, Georgia Institute of Technology}
\affil[2]{Department of Civil and Environmental Engineering, University of California at Davis}
\affil[3]{School of Civil and Environmental Engineering, Georgia Institute of Technology}
\affil[4]{School of Computer Science, Georgia Institute of Technology}

\date{\today}

\begin{document}
\maketitle

\begin{abstract}
\noindent \textbf{Problem definition:}
On-Demand Multimodal Transit Systems (ODMTS) adopt a transit-centric model, integrating fixed-route public transit with on-demand shuttles to provide integrated, door-to-door multimodal service. Although simulation studies have demonstrated the theoretical potential of ODMTS, the system has not yet been tested in practice at scale.
\textbf{Methodology:}
In collaboration with the transit agency of the Atlanta metropolitan region (MARTA), the authors organized \MartaReach{}, a six-month on-demand mobility pilot in Atlanta to explore the practical viability of ODMTS. \Reach{} offered on-demand services across four different zones, connecting each zone to the existing transit network and offering an integrated flat fare.
This allowed for testing key system components in real-world conditions.
\textbf{Results and managerial implications:}
Ridership steadily increased throughout the pilot, suggesting potential for growth if a full ODMTS were implemented.
\Reach{} offered a highly valued service with 94\% satisfaction rate, and 12.24\% of survey respondents indicated they would have driven, ridden with someone, or taken a taxi or ride-hailing service otherwise.
The vast majority of \Reach{} trips were multimodal, especially in West Atlanta, where over 70\% of morning riders connected to rail. The paper presents a potential path to economic viability and discusses practical challenges encountered during the pilot, along with strategies for future deployments.

Following the pilot, MARTA officially adopted \MartaReach{} as a long-term component of its transportation services in March 2026.
The system-wide on-demand transit service currently serves 12 zones with the integrated flat fare and the organizational structure advocated in the pilot.
The adoption decision demonstrates that the insights, system components, and operational lessons documented in this paper translated directly into a larger real-world deployment, providing an opportunity to evaluate ODMTS performance at scale.
\end{abstract}

\noindent \textbf{Keywords:} On-Demand Transit, Microtransit, Multimodal Transit System, Mobility as a Service, Pilot

\section{Introduction}
\label{sect:introduction}
Addressing the first/last mile problem is a critical challenge for cities in the United States. Fixed-route systems struggle to attract riders, especially in low-density areas where providing frequent service is too expensive. This results in riders having to walk long distances to reach their stops, and buses running mostly empty, even in areas where transportation is essential for access to jobs, health services, groceries, and education. Mobility as a Service (MaaS) \citep{shaheen2016mobility} uses modern IT technology to mitigate some of these issues by introducing on-demand micro-transit services. However, these services are not always fully integrated into the existing transit system.

\begin{figure}[!htbp]
	\centering
	\includegraphics[width=0.8\textwidth]{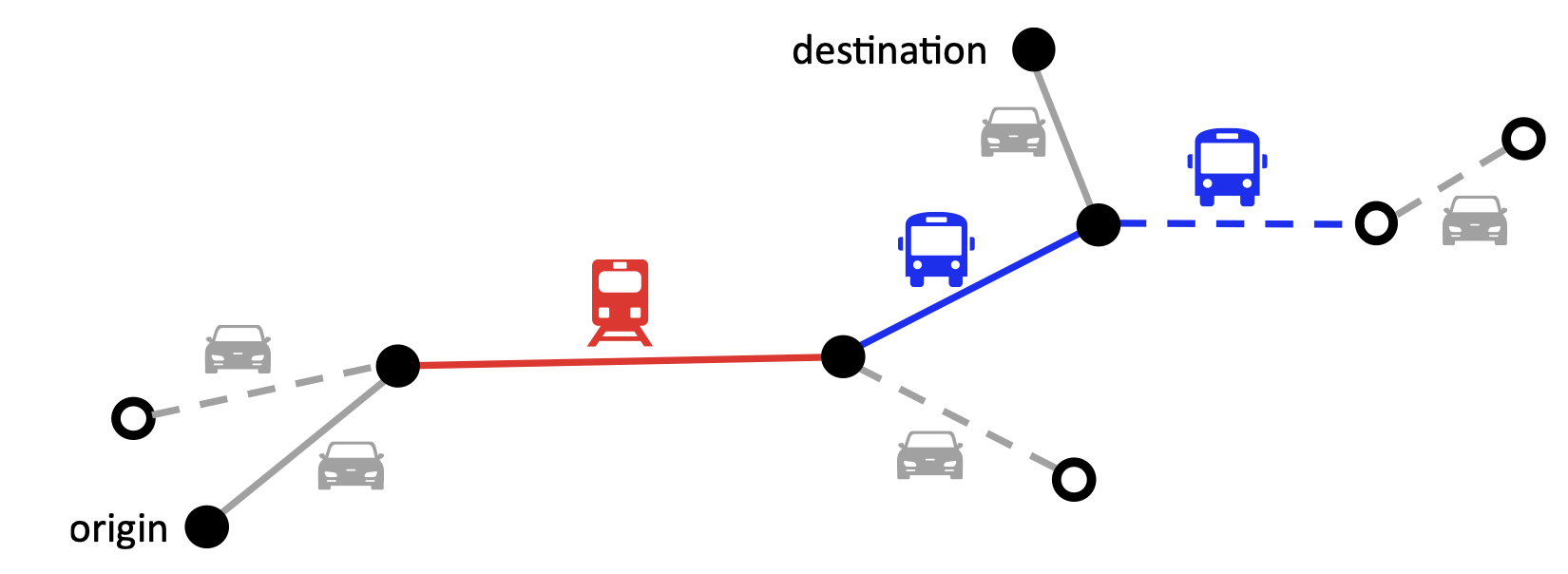}
	\caption{Example ODMTS with Passenger Path (solid lines)}
	\label{fig:ex_ODMTS}
\end{figure}

The concept of On-Demand Multimodal Transit Systems (ODMTS)
\citep{van2019demand} takes a transit-centric approach to
integrate on-demand services and address the first/last mile
problem. ODMTS combine fixed routes and on-demand shuttle services
by design. This integration offers a door-to-door
multimodal service with fully integrated operations and fare
	structure. The on-demand shuttles in ODMTS act as feeders to and
from the fixed rail/bus routes.  In turn, the fixed routes enable
economies of scale and minimize congestion on the high-density
corridors.  Figure~\ref{fig:ex_ODMTS} illustrates the service of ODMTS
for a single rider. Riders provide their origin and destination
through a mobile application or phone call and are offered a route to
serve them. In the figure, a passenger is picked up by an on-demand
shuttle at, or close to, their origin and brought to the train
station. The rider is then instructed to take a train and a bus, which
both run on a fixed schedule. When the rider arrives at the bus
station closest to their destination, another on-demand shuttle is
ready to pick them up and serve the last mile.

ODMTS has received significant attention in recent years, with simulation studies and even a small-scale pilot on the campus
of the University of Michigan.  The case studies cover a wide range of
different cities, including Canberra, Australia
\citep{maheo2019benders}, the Ann-Arbor and Ypsilanti region in
Michigan \citep{van2019demand}, the city of Atlanta, Georgia
\citep{auad2021resiliency}, and Austin, Texas
\citep{lu2023revitalizing}.  A consistent finding throughout these
studies is that ODMTS has the potential to improve cost, convenience,
and accessibility, compared to the existing transit system.
For example, \citet{auad2021resiliency} design and evaluate an ODMTS for the city of Atlanta, a large and complex city in need of a more substantial transit system.
The resulting ODMTS, combining the existing rail network with 24 high-frequency buses and 1100 on-demand shuttles, reduced the average trip duration from 46 to 27 minutes, while simultaneously reducing the cost by over 5\%.

While simulation studies show that ODMTS is theoretically viable, they do not necessarily capture all realities in the field.
A fundamental open question remains: \emph{is ODMTS viable in practice, and does it deliver the benefits observed in simulation studies?}
To make progress toward answering this question, the authors of this paper organized a six-month on-demand mobility pilot known as MARTA Reach (stylized as \Reach{}), which took place in Atlanta in 2022.

\Reach{} provided on-demand services across four zones with distinct characteristics, connecting them to the existing transit network and offering an integrated flat fare regardless of the combination of transit modes used.
The technology was developed by the authors, while the infrastructure was provided by Atlanta’s transit agency.
Although the pilot was not yet a complete realization of ODMTS, which would require significant changes to the transit system, \Reach{} tested key system components and revealed practical insights to guide larger-scale deployment.

Throughout the pilot, data was collected from two main sources: (1) the operational data from the mobile applications that monitored the system, and (2) electronic surveys that took place during and after the pilot.
This paper reports on the results of the pilot and offers insights into the following questions:

\begin{enumerate}
	\item Can ODMTS attract new riders to the system?
	\item Will ODMTS encourage travelers to shift away from cars, taxis, and ride-hailing services?
    \item Are riders willing to make multimodal trips?
    \item Can ODMTS match the service quality observed in simulation studies?
	\item What are the challenges of deploying ODMTS in practice?
    \item Will ODMTS be economically sustainable?
\end{enumerate}

Altogether, this paper contributes a unique perspective on the potential of ODMTS and provides lessons for future implementations.
It also presents substantial evidence that ODMTS may fill an important gap in mobility and deserves further attention.
Lessons about driver behavior, riders gaming the system, and the experience of iteratively designing the applications are aspects that simulation studies alone cannot capture, but that are essential information for ODMTS deployments.

While the pilot contributes substantial evidence toward answering the questions above, some questions can ultimately only be answered through large-scale deployments.
For example, it remains unclear how these findings will evolve as the service grows, particularly regarding rider adoption and mode shift away from cars, service quality under higher demand and more ride-sharing, and the intensity of operational challenges at scale.
Throughout the paper, distinctions will therefore be made between what can be learned from the pilot, and the open questions that remain for full-scale deployment.

\subsection{Adoption of MARTA Reach}
\label{sect:adoption}

The strongest evidence of the genuine value delivered by the \MartaReach{} pilot is MARTA's decision to bring the service back at scale. On March 7, 2026, MARTA officially launched \MartaReach{} as a long-term (five years), system-wide on-demand transit service across 12 zones \citep{MARTA-Reach-website}, spanning communities throughout the Atlanta metro area. Critically, the 2026 service retains the same name (\MartaReach{}), the same integrated flat fare (\$2.50 per trip, matching MARTA's standard bus and rail fare and including free transfers within the system), and the same organizational model under MARTA's operational control. This decision by such major transit agency provides a strong signal: the ODMTS benefits highlighted during the 2022 pilot were convincing enough to scale the services both in space and time.
 This paper therefore serves not only as an account of the pilot, but also as a record of why \MartaReach{} was ultimately adopted by the agency.

\section{Literature Review and Comparison with Existing Services}
\label{sect:literature}

Recent research has shown that the first/last-mile experience significantly impacts transit riders' satisfaction and their loyalty to public transit \citep{venter2020measuring, park2021first}. Using technology-based ride-sharing services to make this connection has become increasingly common. This integration falls under the broader concept of Mobility as a Service (MaaS), which has been widely studied \citep{shaheen2016mobility, feigon2016shared}. As noted by \citet{wang2023economic}, many U.S. transit agencies have expressed strong interest in incorporating on-demand shared mobility into their fixed-route services. Prior work showcases that this integration can boost transit ridership \citep{yan2019integrating}. Studies in low-income communities found that riders preferred integrated services over traditional fixed-route systems \citep{yan2021mobility, wang2022identifying}. Several computational studies have assessed the feasibility and benefits of integrating shared mobility with public transit through simulation and optimization \citep{shen2018integrating, stiglic2018enhancing, maheo2019benders, gurumurthy2020first, imhof2020shared}. These studies typically highlight benefits such as reduced operating costs, improved convenience, and solutions to first/last mile gaps.

Actual on-demand services have also been implemented to bring riders to transit services. However, these projects are often operated as independent services. Some pilot operators include {\sc RideCo}, a technology company that launched a one-year pilot in March 2015 with the transit agency in Milton, Toronto, Canada
\citep{rideco2018solving}. In 2017, MaaS company Via started to
initiate partnerships with cities in Europe and in the United States
\citep{via2023}. In September 2018, the city of Belleville in Canada
ran an on-demand transit pilot for six months, which substituted a
late-night fixed-route with the on-demand service
\citep{sanaullah2021spatio}. However, the on-demand service was only
operated between existing fixed-route stops; hence, this pilot did not
aim to provide connections to fixed-routes. Other mobility on-demand
pilots have been carried out in different locations in the world such
as Dhaka, Bangladesh \citep{kamau2016demand}, Munich, Germany
\citep{mvg2018service}, and Milpitas, California,
United States \citep{milpitas2022}.

Several recent pilots in the United States were supported by the
Mobility on Demand (MOD) Sandbox Program to address the first/last
mile problem. The MOD Sandbox Program is an initiative with the goal
of promoting and advancing innovative transportation solutions within
the United States \citep{patel2022mobility,fta2023}. Led by the
U.S. Department of Transportation (DOT), the program aims to
facilitate collaborations between the public and private sectors to
tackle mobility challenges and enhance transportation options for
communities \citep{fta2023}. In Pierce County, Washington, a pilot was conducted in
collaboration with the ride-hailing company Lyft and transit
agencies. The pilot targeted individuals who required transportation
beyond half a mile from the nearest transit access point, or who
needed service after regular transit hours. Its objectives were to
improve access to transit bus routes, increase transit ridership, and
reduce congestion \citep{cordahi2018mod_pierce,
	brown2022mobility_pierce}. In Los Angeles County, California, and
King County, Washington, two pilots were implemented in collaboration
with multiple transit agencies and Via. These programs aimed to
provide equitable first/last mile transit access to fixed-routes,
allowing riders to request Via rides to/from transit stations within
specified zones \citep{cordahi2018mob_la,
	martin2022mobility_la}. Moreover, a pilot in Dallas, Texas, explored
a soft integration of smart app platforms for Transportation Network
Companies (TNCs) and other MOD providers like Spare. This pilot aimed
not only at providing first/last mile solutions to DART (Dallas Area
Rapid Transit) riders but also at assessing the potential of replacing
ineffective fixed-route services in low-density areas with on-demand
mobility services \citep{cordahi2018mod_dallas,
	martin2021mobility_dallas}.

The pilots and studies above established collaborations between transit agencies and TNCs, or were solely led by the TNCs themselves. ODMTS is fundamentally different:  \emph{it envisions the future of transit systems to be completely integrated and run entirely by transit agencies.} This integration allows for holistic design and operations, a simple fare structure, and economies of scale.
The \Reach{} pilot serves as a prototype of the ODMTS concept, which has been studied extensively in the literature.
Appendix~\ref{sect:odmts_additional} summarizes key technical advances related to ODMTS, several of which informed the design of \Reach{}.
In particular, the pilot draws inspiration from the pipeline proposed by \citet{auad2021resiliency}, and incorporates both the real-time shuttle dispatcher developed by \citet{riley2019column} and the shuttle
rebalancing algorithm introduced by \citet{riley2022operating} and
further studied by \citet{lu2026impact}.

Table~\ref{tab:compare_table} highlights some of the key differences between connected services offered by TNCs, the \MartaReach{} pilot, and a fully realized ODMTS.
In both \Reach{} and ODMTS, the benefits and revenues remain within the public transit agency.
In contrast, partnerships with TNCs involve two separate transportation providers, which can result in conflicting interests and uneven access to information.
\Reach{} and ODMTS also allow agencies to manage their own drivers, making it easier to maintain workforce stability when transitioning to a new transit model.
For example, if certain fixed bus routes are reduced or restructured in favor of on-demand shuttles, the same pool of drivers can be reassigned to operate these shuttles.
Unlike TNC-based connections, where fares are determined by the private operator and may include surge pricing, \Reach{} and ODMTS charge an integrated flat fare consistent with the broader transit network, ensuring equitable access regardless of trip distance or time of day.

\begin{table}[!t]
	\centering
	\caption{Key Characteristics of TNC Connections, MARTA Reach, and ODMTS.}
	\label{tab:compare_table}
	\resizebox{1\textwidth}{!}{
		\begin{tabular}{l l l l l l l l}
			\toprule
			\textbf{Service} & \textbf{Business Model} & \textbf{Operates On} & \textbf{Fleets Ownership} & \textbf{Drivers Employment} & \textbf{Multimodal Trips} & \textbf{Fares} \\ \midrule
			TNC Connections & \makecell[l]{TNC \& agency \\ collaboration} & \makecell[l]{Ride-hailing vehicles} & TNC or individuals & \makecell[l]{Contracts with TNC} & \makecell[l]{Rider selects \\ transfer stops} & \makecell[l]{Determined by TNC} \\ \midrule
			\makecell[l]{\MartaReach{}} & \makecell[l]{Agency-operated \\ platform} & \makecell[l]{On-demand shuttles} & Transit agency & \makecell[l]{Managed by agency} & \makecell[l]{Rider selects \\ transfer stops} & \makecell[l]{Integrated fare with \\ transfers to fixed-routes} \\ \midrule
			ODMTS & \makecell[l]{Agency-operated \\ platform} & \makecell[l]{On-demand shuttles \\ \& fixed routes} & Transit agency & \makecell[l]{Managed by agency} & \makecell[l]{System provides \\ full trip-plan} & \makecell[l]{Integrated fare with \\ transfers to fixed-routes} \\ \bottomrule
		\end{tabular}
	}
\end{table}

ODMTS is not an isolated research topic, as it has the potential to integrate with other ongoing research efforts in public transit.
These include fleet electrification \citep{holland2021environmental, abdelwahed2021boost}, autonomous vehicle operations \citep{zardini2022analysis, carreyre2024demand}, paratransit system integration \citep{collins2025landscape, chen2025empowering}, and improved accessibility for riders with special needs \citep{de2025optimizing}.
Some of these integrations have already been considered for \MartaReach{} and later pilots.
Further discussion of these topics is provided in the following sections, including Section~\ref{sec:followup}.

\section{Methods: Pilot Design}
\label{sect:design}

The \Reach{} pilot was conceived to evaluate the ODMTS concept in a complex city. Atlanta is the 10th most congested city in the U.S. \citep{Pishue2023GlobalTrafficScorecard} and showcases many characteristics that impose challenges when deploying an on-demand pilot. The riders, the transit
system, and the realities in the field interact to
raise these challenges. In particular, the population and the
neighborhoods are diverse, with different socio-economic backgrounds,
accessibility needs, mobility options, and perceptions of the transit
system. Some neighborhoods are residential, some are job centers, and
some are a combination of both. MARTA, the transit system of the
Atlanta area, offers multiple modes of transportation including rail, bus, and paratransit services; some of these are directly operated by MARTA while some others are contracted, bringing other financial and integration challenges.

This section presents a comprehensive overview of the \MartaReach{}
pilot: it presents the neighborhoods selected for
\Reach{} and highlights the concepts behind its operations. \Reach{}
operated from March~1, 2022 to August~31, 2022 from 6:00 a.m.\ to 7:00
p.m.\ on weekdays, including federal holidays, for a total duration of
132 service days.
The pilot was deployed in two phases that are detailed in Appendix~\ref{sec:timeline}. On each day, two shuttles were kept in reserve, while 16 shuttles were distributed over the pilot zones. Each \Reach{} shuttle has a capacity of eight people and is wheelchair accessible.

\Reach{} was a collaboration between the Socially Aware Mobility Lab at Georgia Tech, the
Metropolitan Atlanta Regional Transit Authority (MARTA), and Propel
Atlanta, a transportation advocacy group.  It was supported by a
National Science Foundation CIVIC grant that covered the development
of the technology and the actual operations.
The Socially Aware Mobility Lab developed
the software technology enabling the pilot, including the mobility and
supervision applications and the back-end computing server.
The overall pipeline was inspired by \citet{auad2021resiliency} and shuttles were dispatched with the algorithm by \citet{riley2019column}.
MARTA
operated the pilot, and provided, through contractors, the shuttle
fleet, the drivers, and the supervisors. The fare for riders was set
equal to the \$2.50 fare for the existing system, whether the rider
used on-demand shuttles, buses, rail, or a combination of all
three. The pilot was deployed in four different zones with distinct
characteristics to understand the potential benefits of ODMTS from a broad
perspective.

\subsection{Planning The Pilot Zones}
\label{subsect:zone}

\begin{figure}[t]
	\centering
	\includegraphics[width=0.6\textwidth]{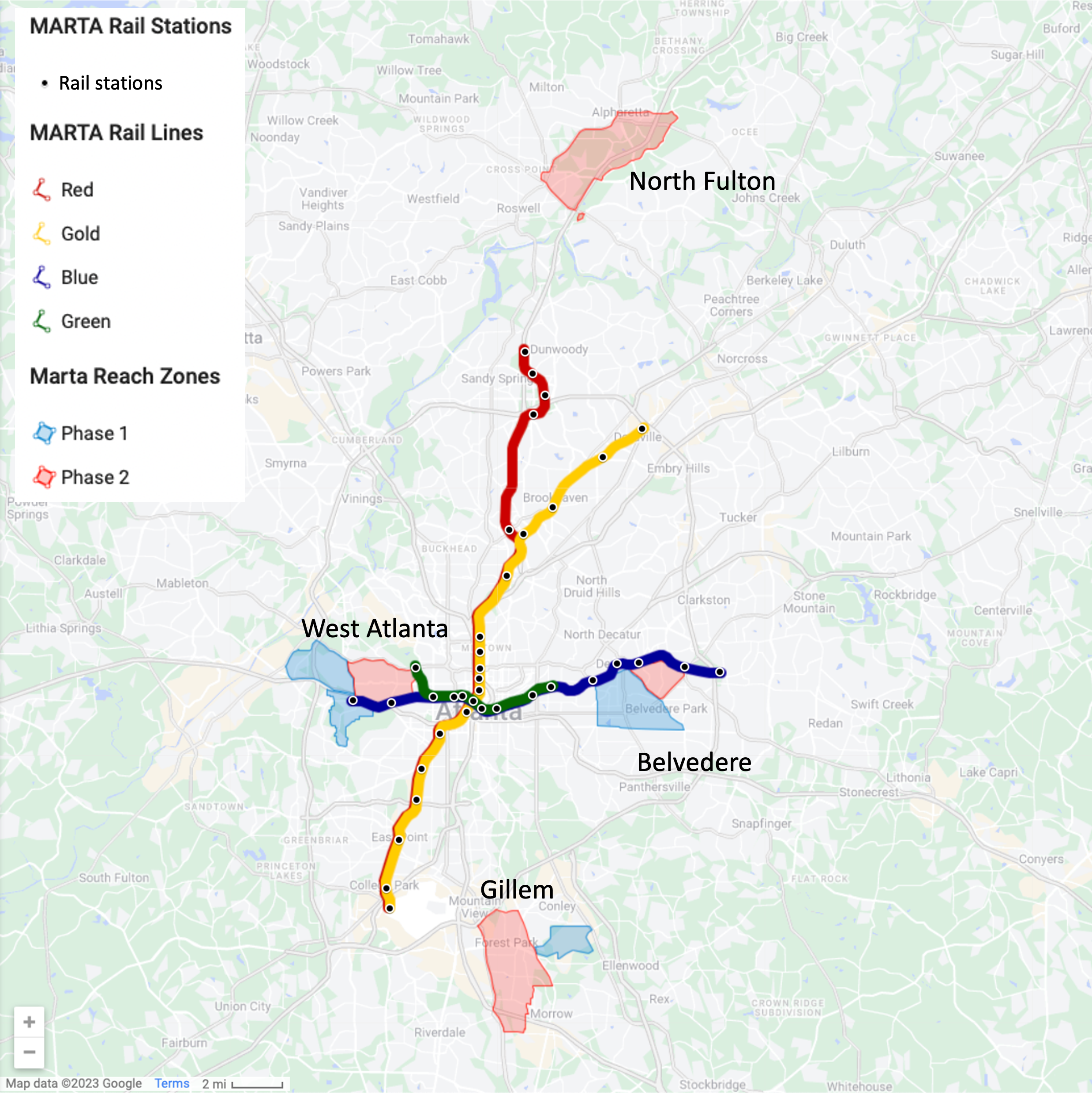}
	\caption{The MARTA Reach Pilot Zones. An interactive map is
	available on
	\href{https://www.google.com/maps/d/viewer?mid=1_pWHpPQSqbI75mEdspi8dUeAGOg5Hdk&usp=sharing}{this
		web site.}}
	\label{fig:all_zones}
\end{figure}

Figure~\ref{fig:all_zones} provides an overview of the Phase~1 and Phase~2 service areas. It also shows how the four zones interact with the existing MARTA rail lines. Each of the selected zones had a potential connection to the overall MARTA system (e.g. by bus), but also had a significant first/last mile problem. The areas were selected to represent different types of neighborhoods. For example, West Atlanta is a low-income residential area, Gillem is a job center, and Belvedere is a mixed-use zone with diverse population. Nearly a year before the launch of \Reach{}, the development teams began exploring potential pilot areas that could benefit local communities and advance transportation research. The final selection of pilot zones was ultimately made by MARTA. Table~\ref{tab:Demographics} provides demographics information for the service areas in both phases, based on data from the 2020 census \citep{census2020} and the 2021 American Community Survey \citep{census2021}, which was the available data when the zones were selected. Although the zones could not be qualified as transit deserts, transit options were quite limited.

\begin{table*}[t]
\caption{Demographics and Shuttle Counts for each Service Area.}
\normalsize
    \begin{subtable}{\textwidth}
    \caption{Phase 1}
	\resizebox{\textwidth}{!}{%
		\centering
		\begin{tabular}{l r r r r r r r r r}
			\toprule
			& & & \multicolumn{4}{c}{\textbf{Race (\%)}} & \textbf{Ethnicity (\%)} & \\
			\cmidrule(lr){4-7} \cmidrule(lr){8-8}
			\multirow{2}*{\textbf{Zone}} & \multirow{2}*{\makecell{\textbf{Area} \\ \textbf{($\text{mi}^2$)}}} & \multirow{2}*{\textbf{Population}}& \textbf{White} & \makecell{\textbf{Black or} \\ \textbf{African} \\ \textbf{American}} & \textbf{Asian} & \textbf{Other} & \makecell{\textbf{Hispanic or} \\ \textbf{Latino}} & \makecell{\textbf{Average} \\ \textbf{per capita} \\ \textbf{income (\$) }} & \makecell{\textbf{Number of} \\ \textbf{shuttles}}\\
			\midrule
			West Atlanta &	4.99 &	11,712 &	3.20 &	93.25 &	0.48 &	3.07 &	3.96 & 26,784.13 & 10\\
			Belvedere &	4.99 &	18,971 &	58.21 &	34.15 &	5.21 &	2.43 &	5.19 & 50,453.83 & 3\\
			Gillem &	1.87 &	2,093 &	10.05 &	63.68 &	2.84 &	23.44 &	23.88 & 22,636.28 & 3\\
			\bottomrule\\
	   \end{tabular}
    }
    \end{subtable}
    \begin{subtable}{\textwidth}
    \caption{Phase 2}
	\resizebox{\textwidth}{!}{%
		\centering
		\begin{tabular}{l r r r r r r r r r}
			\toprule
			& & & \multicolumn{4}{c}{\textbf{Race (\%)}} & \textbf{Ethnicity (\%)} & \\
			\cmidrule(lr){4-7} \cmidrule(lr){8-8}
			\multirow{2}*{\textbf{Zone}} & \multirow{2}*{\makecell{\textbf{Area} \\ \textbf{($\text{mi}^2$)}}} & \multirow{2}*{\textbf{Population}}& \textbf{White} & \makecell{\textbf{Black or} \\ \textbf{African} \\ \textbf{American}} & \textbf{Asian} & \textbf{Other} & \makecell{\textbf{Hispanic or} \\ \textbf{Latino}} & \makecell{\textbf{Average} \\ \textbf{per capita} \\ \textbf{income (\$)}} & \makecell{\textbf{Number of} \\ \textbf{shuttles}}\\
			\midrule
			West Atlanta &	8.42 &	21,031 &	4.39 &	92.64 &	0.49 &	2.48 &	3.47 &	22,303.73 & 6 \\
			Belvedere &	6.79 &	24,816 &	59.56 &	33.01 &	4.99 &	2.44 &	5.06 &	51,208.51 & 4 \\
			Gillem &	9.39 &	20,264 &	13.00 &	46.83 &	8.93 &	31.24 &	31.85 & 19,435.61 & 	2 \\
			North Fulton &	8.52 &	23,012 &	39.41 &	18.76 &	11.7 &	30.13 &	30.83 &	40,071.29 & 4 \\
			\bottomrule
	\end{tabular}
    }
    \end{subtable}
\label{tab:Demographics}
\end{table*}

\subsubsection{West Atlanta}
The West Atlanta zone is a residential area with relatively low average income compared to the overall Metro Atlanta income of \$39,904 per capita \citep{census2021}.
The key mobility assets in this zone are the H.E. Holmes, Westlake, and Bankhead rail stations which connect the West-Atlanta neighborhoods to the blue and green lines of the MARTA rail system.
However, the overall zone is large and has limited bus coverage to reach these train stations.
In terms of demographics, Tables~\ref{tab:Demographics} show that the population of West Atlanta is predominantly Black or African American.

In Phase~1, ten \MartaReach{} shuttles operated in the blue area to serve residents in five middle and lower income neighborhoods: Fairburn Heights, Westhaven, Collier Heights, Peyton Forest, and Florida Hill \citep{incomeAtlanta}. The expansion in Phase~2 added five more neighborhoods: Center Hill, Dixie Hill, Penelope Neighbors, West Lake, the southern half of Grove Park, and the southern third of Carey Park. Although the expansion increased the area and population served, only six shuttles operated in this zone during Phase~2. The other four shuttles were relocated to the new North Fulton zone.

\subsubsection{Belvedere} Belvedere zone locates at the east of Atlanta around the Belvedere Park. This zone is a mixed-use neighborhood with residential and commercial activities. There are some significant differences in income levels between various parts of the zone. More precisely, residents are  distributed along five main high and middle-income neighborhoods: East Lake, Oakhurst, College Heights, Midway Woods, and White Oak Hills \citep{incomeAtlanta}. The commercial area includes the Walmart Supercenter in the southern section of the red area in Figure~\ref{fig:all_zones}: this is both a job center and a shopping center. The Belvedere zone is connected to the existing MARTA rail blue line through the East Lake, Decatur, and Avondale stations in the northern part of the zone. There are also buses running through the south part of the zone. Table~\ref{tab:Demographics} show that its population is predominantly White with a strong share of Black or African American residents, and some Asian population.
Phase~2 added two more neighborhoods, Avondale Estates and the east of Belvedere Park, and an additional shuttle was introduced.

\subsubsection{Gillem} The Gillem zone locates south of Atlanta near the Hartsfield-Jackson International Airport. This zone was chosen mainly because it is an industrial zone that has no direct connectivity to the existing MARTA rail system. In Phase~1, three shuttles operated in the blue area of this zone, focusing primarily on serving the Gillem Logistics Center with facilities such as the Kroger Fulfillment and Distribution Centers. In Phase~2, the Forest Park neighborhood was added.
This zone is quite diverse, with a majority of the population being Black or African American, complemented by a strong Hispanic or Latino representation. Although the expansion increased the area and population served, only two shuttles operated in this zone during Phase~2. The third shuttle that was originally operating in this zone was moved to serve the Belvedere zone during Phase~2.

\subsubsection{North Fulton} The North Fulton zone is located north of Atlanta near the city of Alpharetta. This zone was included because it is very diverse in race, ethnicity, and income; it is a mixed-use region that joins residential and commercial areas and has no direct access to the existing MARTA rail lines.

\subsection{Important Designs for Pilot Operations}
\label{subsect:pilot_operation}

This section reviews some practical aspects to operating \MartaReach{}.
This does not only provide guidance for future implementations, but also provides context to understand the findings in the next section. Rider outreach and community awareness campaigns were managed by MARTA, which advertised the service through branded vehicles, station signage, and targeted community outreach in each pilot zone.

\subsubsection{Virtual Stops Selection}
The \MartaReach{} pilot used the concept of \emph{virtual stops},
i.e., virtual locations on a map where shuttles pick up or drop off passengers.
Virtual stops make it easier for drivers and passengers to find each other, and it is a practical mechanism to exclude unsafe or inconvenient drop-off points.
Because there is no need for a marker in the real world, it is easy to add or remove virtual stops, and many virtual stops were used to ensure that there is always a stop nearby, making the pilot as close to ``door-to-door'' as possible.
Figure~\ref{fig:virtual_stops} shows the Belvedere zone as an example.
Virtual stops were located at MARTA bus stops and rail stations (blue) to facilitate easy transfers, and additional stops where created throughout the zone (orange).
The development team automatically generated candidate virtual stops to ensure coverage within approximately 50 yards of walking distance, after which MARTA's safety team manually reviewed and approved each stop to ensure safety and accessibility.
The flexibility of virtual stops made it possible to evaluate and update the locations on a weekly basis, based on how often each stop was used.
At the end of the pilot, 1436 virtual stops were in use over all zones, capturing 951 existing MARTA stops and 485 additional locations. Of the computationally generated candidate stops, only 45\% (not including existing MARTA stops) were retained after the manual review. In the final selection, the average distance between each stop and its nearest neighboring \Reach{} stop is approximately 164 yards (150.20 meters), normally under a two-minute walk, as computed using Graphhopper on OpenStreetMap data \citep{graphhopper2026}.

\subsubsection{Idle Locations Preparation}
\begin{figure}[htbp]
	\centering
	\includegraphics[width=\textwidth]{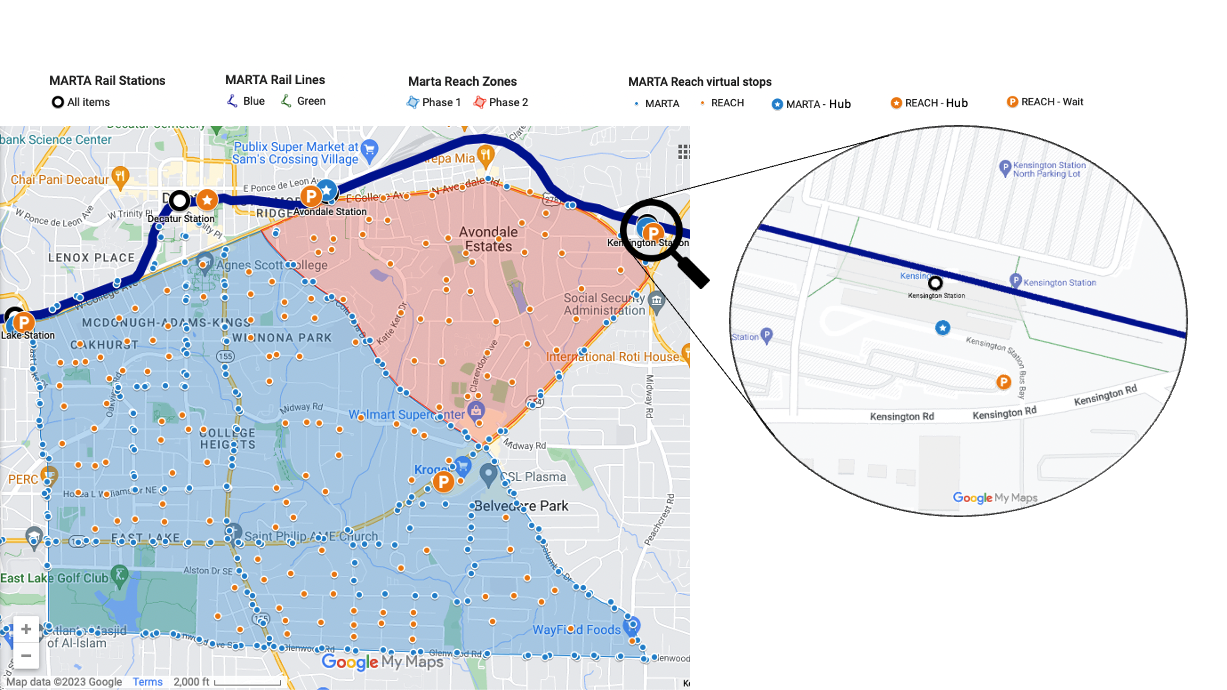}
	\caption{The MARTA Reach Virtual Stops in Belvedere.}
	\label{fig:virtual_stops}
\end{figure}

Another practical concern is that not all locations are safe for shuttles to wait for trip assignments. This issue was addressed by introducing \emph{idle stops} where idling was allowed. If there were no outstanding requests after completing a trip, the driver mobile application would route the shuttle to the nearest idle location to wait for the next request. The idle stops were chosen by the MARTA safety team from a list of all the virtual stops.
The safety team was very conservative in their decisions, mostly approving rail stations and authorized parking lots. The zoom-in in Figure~\ref{fig:virtual_stops} shows the idling stop at Kensington Station in Belvedere as an example. In total, there were three idling stops in West Atlanta, and four each in Belvedere, Gillem, and North Fulton. In Phase 1 of the pilot, vehicles were typically directed to the nearest idle stop after completing a trip.

In Phase 2, a new rebalancing algorithm was introduced to reposition vehicles toward areas with higher expected demand, as detailed by \citet{lu2026impact}. The system collects recent trip requests near each idle stop to estimate where demand is emerging. It then solves an optimization problem that minimizes the ratio of recent requests to assigned vehicles across all idle stops, so that the number of idle shuttles is proportional to demand. Importantly, this solution only determines how many vehicles should be assigned to each idle stop; a second optimization model then takes these target counts as input and assigns specific idle vehicles to specific stops, minimizing total repositioning travel time across the fleet. Vehicles already parked at an idle stop are exempted to avoid unnecessary movement. The trade-off is that repositioning adds some empty driving, but the benefit is a better-distributed fleet that reduces rider wait times. A key advantage of this approach is that it does not require extensive historical ridership data, making it well-suited for early-stage services where data is still limited. For longer-term deployments where sufficient ridership data has accumulated, machine learning techniques can be integrated to further improve demand prediction and guide repositioning \citep{riley2020real}. The formal mathematical formulation of the rebalancing algorithm is provided in Appendix~\ref{sec:opt_models}.

\subsubsection{Shuttle Dispatcher}
The \Reach{} dispatcher uses the column generation algorithm introduced by \citet{riley2019column}. Each pilot zone is optimized independently as shuttles do not cross zones. Rather than sending each new rider request to the nearest available vehicle, the algorithm batches all pending requests at the end of each short time epoch (e.g., 30 seconds) and solves a global routing optimization problem. Specifically, the algorithm first generates candidate routes for each vehicle, i.e., sequences of pickup and drop-off stops for riders that also respect vehicle capacity and travel-time deviation limits. The optimization model then finds the combination of routes that minimizes total rider wait time across the entire fleet.The key idea is that jointly optimizing all requests at once allows the system to pair riders heading in similar directions, reserve closer vehicles for riders who will benefit most, and avoid locally greedy decisions that create longer waiting times for other riders. By contrast, a naive nearest-vehicle rule minimizes the distance for one rider at a time without considering the full pool of pending requests or the potential for ride-sharing. The formal mathematical formulation of the optimization model for the dispatcher is provided in Appendix~\ref{sec:opt_models}.

\subsubsection{Request Interfaces and Fare Program Designs}
Riders had the option to make and cancel trip requests either through the \Reach{} mobile application or via a phone call. \Reach{} only provided service between origins and destinations in the same pilot zone, but thereby enabled transfers to fixed-route service that continue outside of the zone. Riders could submit a request for up to four people so groups could travel together. In the event that riders failed to appear at the
pick-up location, the trip was recorded as a \textit{no-show} and considered a cancellation. A cancellation during this pilot did not result in any penalty for the rider. Details about the technologies behind the mobile applications are provided in Appendix~\ref{sect:technology}.

\Reach{} riders were charged the traditional MARTA flat fare of \$2.50, including transfers to fixed-route services. Payment could be made using cash or MARTA Breeze cards. The flat rate paid for the entire multimodal trip, like the one depicted by the solid line in Figure~\ref{fig:ex_ODMTS}. In the \MartaReach{} pilot, riders were responsible for planning the \Reach{} connections that they needed to complete their multimodal trip.

\subsubsection{Follow-up Development after Reach}
\label{sec:followup}
A possible improvement for future \Reach{}-like pilots is to offer automatic multimodal trip planning to simplify travel for riders.
This feature was implemented by the authors in a 2024-2025 follow-up pilot in Savannah, GA \citep{cat2025}, marking another step toward a fully integrated ODMTS.
In addition, instead of relying on virtual stops, the Savannah deployment allowed riders to request door-to-door service by selecting any origin and destination within the service area.
The vehicles were also downsized from large vans to smaller electric vehicles that maintain wheelchair accessibility while improving operational efficiency.
A full analysis of the Savannah pilot is planned to be reported in a future paper.

\section{Results: Pilot Evaluation and Economic Analysis}
\label{sect:results}

This section presents the evaluation of the \MartaReach{} pilot from multiple perspectives.
The results in this section are based on real-time operational data that was collected for all users through the mobile applications.  In addition, the operational data was complemented with rider surveys. A long-term evaluation survey was distributed to \Reach{} users via email on August 8, 2022, and 232 responses (before data cleaning) were collected until September 23, 2022.  Trip-specific surveys were available to the riders throughout the entire pilot period, collecting 262 total responses (before data cleaning). A summary of the survey results is provided here, and the reader is referred to \citep{drake2024evaluation} for a comprehensive analysis of the travel behavior of \Reach{} riders.

\subsection{Ridership}
\label{subsect:ridership}

\begin{figure}[!p]
	\centering
	\begin{subfigure}{\textwidth}
		\includegraphics[trim={0 1.4cm 0 2cm},clip,width=\textwidth]{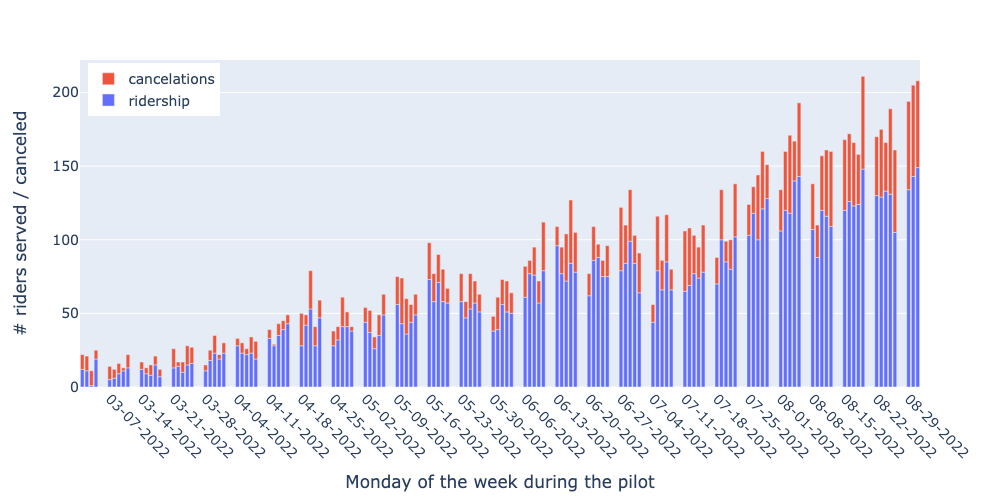}
		\caption{Daily.}
		\label{fig:ridership_daily}
	\end{subfigure}
	\begin{subfigure}{0.55\textwidth}
		\includegraphics[trim={0 1.4cm 0 2cm},clip,width=\textwidth]{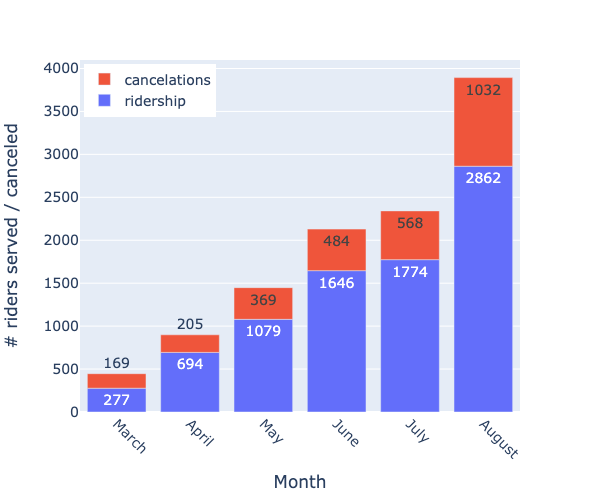}
		\caption{Grouped by Month.}
		\label{fig:ridership_monthly}
	\end{subfigure}
	\begin{subfigure}{\textwidth}
		\includegraphics[trim={0 1.4cm 0 2cm},clip,width=\textwidth]{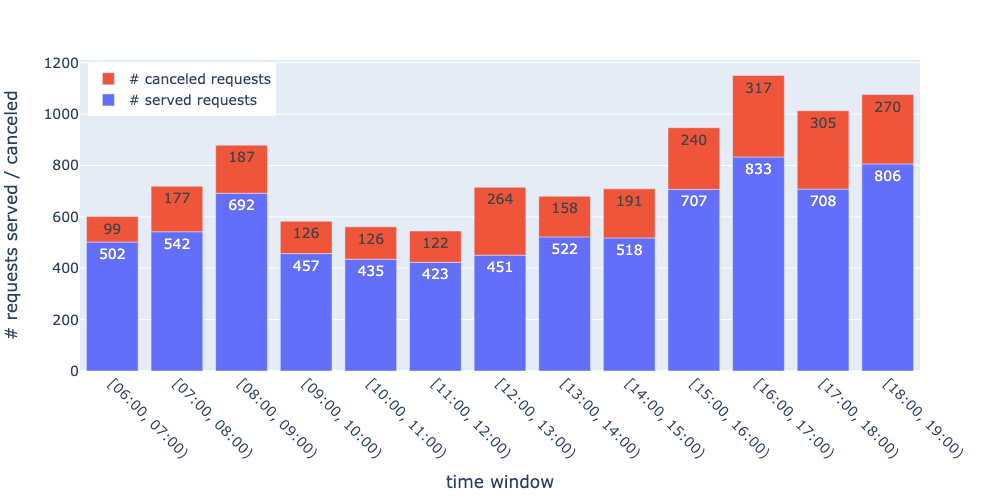}
		\caption{Grouped by Time Period.}
		\label{fig:request_time_distribution}
	\end{subfigure}
	\caption{MARTA Reach Ridership and Cancellations During the Pilot Period.}
	\label{fig:ridership}
\end{figure}

Figure~\ref{fig:ridership} shows how \MartaReach{} ridership evolved over the pilot period.
Ridership was continuously increasing, and ridership increased by over 61\% in the last month.
This shows potential for continued growth if a full ODMTS were implemented, but also highlights that there may be a lag between introduction of the new system and rider adoption.
Throughout the pilot period, \Reach{} served 7,596 requests, corresponding to a
total of 8,332 riders. Ridership was relatively low on federal
holidays, i.e., Memorial Day (May 30, 2022), Juneteenth (observed June
20, 2022), and Independence Day (July 4,
2022).
West Atlanta and Belvedere accounted for the majority of the
rides, with 58.32\% and 34.97\%, respectively;
Gillem and North Fulton served only 3.90\% and 2.80\% of the rides.
This difference is likely due to the fact that Gillem and North Fulton do not connect to any rail stations, which will be explored in Section~\ref{subsect:case_for_multimodal}.
Another interesting observation in Figure~\ref{fig:ridership} is the high number of cancellations, many of which were followed by an actual ride soon after. Section~\ref{sect:cancel} will dive into this behavior in more detail.

Figure~\ref{fig:request_time_distribution} reports the distribution of
trip requests over the service hours: demand was generally higher during
peak hours, particularly between 8:00 to 9:00 a.m.\ and after 3:00 p.m.  The
majority of requests (92.33\%) were submitted by individual riders. A
small portion of requests were submitted for groups of two, three, or
four riders (470, 70, and 42 requests, respectively).
Shuttles were occasionally assigned multiple requests at the same time, particularly when demand was high.
Ride sharing accounted for
9.66\% of passenger-serving mileage.

As expected for an early-stage pilot, ride sharing was limited and vehicles rarely operated near capacity. A full-scale ODMTS deployment would involve higher demand, more frequent ride sharing, and longer detours, and understanding how service quality and rider behavior evolve under those conditions is a natural and important next step.
Prior simulation studies suggest that even at metropolitan scales, ODMTS can deliver efficient service with minimal ride sharing \citep{auad2021resiliency}, and the limited ride sharing observed in this pilot is consistent with these computational expectations. Confirming whether this holds under full operating conditions remains an important open question, and the 2026 \MartaReach{} service, now running at a substantially larger scale within MARTA services, is well positioned to answer that question.
The \MartaReach{} pilot operated eight-seat paratransit vehicles that were already part of the MARTA fleet.
This choice was practical for the pilot, but given the limited ride sharing observed and in alignment with \citet{auad2021resiliency}, smaller four-seat sedans may be more suitable for future implementations.

\begin{figure}[htbp]
	\centering
	\includegraphics[trim={0 0 0 2.5cm},clip,width=0.55\linewidth] {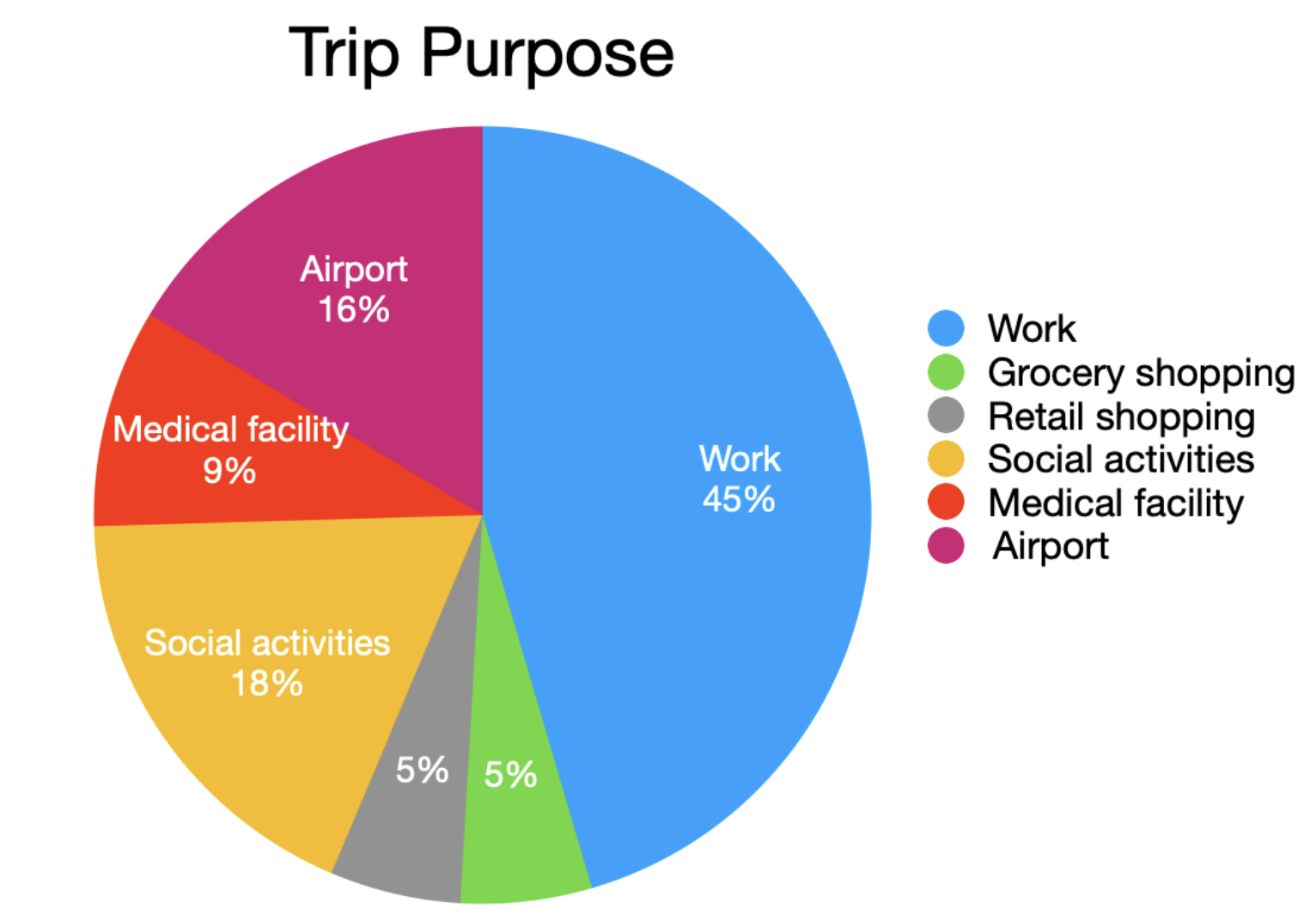}
	\caption{Breakdown of Popular Trip Purposes.}
	\label{fig:trippurpose}
\end{figure}

When asked about their experience, 94\% of the riders report to be {\em satisfied} or {\em very satisfied} with \MartaReach{}.
Another item in the survey asks the same question for MARTA services overall, and reveals that, for these riders, \Reach{} provided a significantly better experience.
For example, 71\% of riders are {\em very satisfied} with \Reach{}, while 35\% of riders are {\em very satisfied} with the traditional MARTA services.
Figure~\ref{fig:trippurpose} shows for what purpose riders use \Reach{}, as obtained from the surveys.
Commuting represents the largest category, making up nearly half of all trips.
Social activities and shopping together make up a large fraction as well, followed by multimodal trips to the airport and trips to access medical care.

\subsubsection{Mode Switching}
\label{sec:mode_switching}
The survey question ``{\em What alternate modes of transportation
	would you have used, had Reach not been available}'' provides
some interesting insights into mode switching behavior.
Survey takers could provide multiple answers, and Table~\ref{tab:mode_switch_stats_hierarchical} summarizes the results with a hierarchical classification into five categories.
That is, if any of the answers falls in the \emph{transit} category, it is conservatively listed in this category.
If any of the answers falls in the \emph{auto} category and none fall in the \emph{transit} category, it is listed in the \emph{auto} category, etc.

The hierarchical classification reveals that most \Reach{} riders were already transit users.
However, 35\% of \Reach{} trips would not have been transit trips otherwise. What stands out is that 12.24\% of respondents indicated they would have driven, ridden with someone, or taken a taxi or ride-hailing service, suggesting that ODMTS can encourage some mode shift away from car-based alternatives.

\begin{table}[!t]
	\centering
	\resizebox{\textwidth}{!}{
		\begin{tabular}{c l l r}
			\toprule
			Rank &  Category  & Travel Modes in the Category & \# Responses (\%) \\
			\midrule
			1 & Transit & MARTA Bus, Rail, and Mobility & 158 (64.49\%) \\
			2 & Auto & Drive myself, Ride with someone, Taxi/Uber/Lyft & 30 (12.24\%) \\
			3 &  Active & Walk, E-Scooter, Bike & 39 (15.92\%) \\
			4 & Other & Others & 2 (0.82\%)\\
			5 & Would not make the trip& N/A & 16 (6.53\%) \\
			\bottomrule
		\end{tabular}
	}
	\caption{Hierarchical Classification of Alternative Travel Modes.}
	\label{tab:mode_switch_stats_hierarchical}
\end{table}

\subsubsection{Frequent Riders}
\label{subsubsect:frequent_riders}
The data shows that \Reach{} attracted some loyal customers that relied on the service daily.
During the
pilot period, over 550 individuals were served by \Reach{}.
A total of 14 riders
took more than 100 trips each and 41 riders used the service more
than 50 times. These Top-14 riders accounted for 2,143 trips (28.21\%
overall), while the Top-41 riders contributed more than half of the
total served requests (3,924; 51.66\%).
One of the riders made 26
requests by calling the dispatch center, showing that \Reach{}
was able to serve people with no access to modern technologies.
In total there were 136 {\em frequent trips}: origin-destination pairs
that were requested more than 10 times by one rider.
As expected for an early-stage pilot, this concentration of frequent users is a natural feature of early adoption. At the same time, it is worth acknowledging that aggregate ridership statistics were largely driven by this small group, and a larger and longer deployment will tell whether the ridership base increases with community awareness.

As an example, one rider who lived in a pilot zone regularly commuted using \MartaReach{}.
Starting March 17, 2022, they used \Reach{} as part of their commute on 74 out of the 119 remaining service days (62.2\%).
They completed 56 full commuting tours, and only took home-to-work and work-to-home trips on 15 and 3 days, respectively.
The identical origin-destination pair for this commute was the most requested trip, with 130 out of 146 requests.
The average \Reach{} travel time to work for this rider was 20.86 minutes (including 9.83 minutes of waiting), while the fixed-route travel time around 8:00 a.m.\ on weekdays is approximately 30 minutes, including walking. Similarly, the work-to-home trips had an average time of 20.81 minutes (including 7.60 minutes waiting time), while the fixed routes take about 25 minutes around 6:00 p.m.\ For this particular rider, \Reach{} offered a shorter travel time for the same fare and eliminated the need for walking the first and last leg.
The data has multiple examples of individuals with similar travel patterns, providing supporting evidence that ODMTS can offer reliable and cost-effective service for local communities.

\subsection{Multimodality and the First/Last Mile Problem}
\label{subsect:case_for_multimodal}

\begin{figure}[htbp]
	\centering
\begin{subfigure}[!t]{0.9\textwidth}
	\includegraphics[trim={0 1.4cm 0 0}, clip, width=\textwidth]{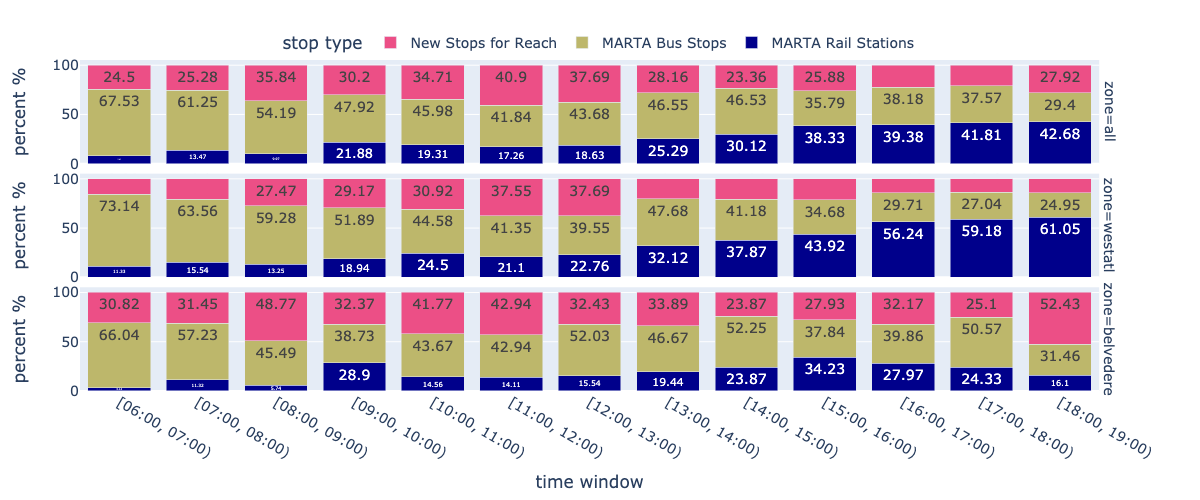}
	\caption{Depart from Origin Stops}
	\label{fig:request_stop_type_as_origin}
\end{subfigure}
\par
\begin{subfigure}[!t]{0.9\textwidth}
	\centering
	\includegraphics[trim={0 1.4cm 0 0}, clip, width=\textwidth]{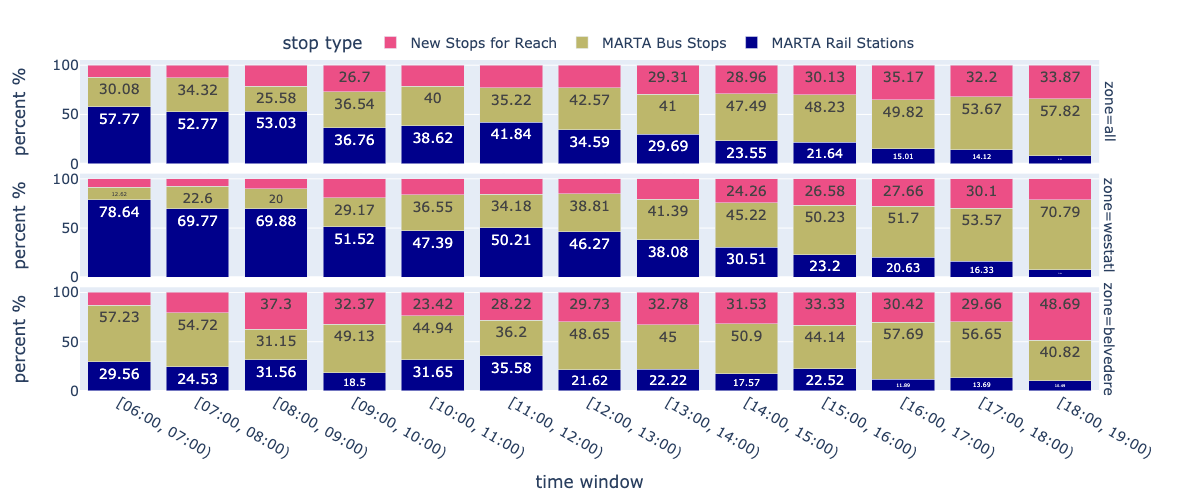}
	\caption{Arrive at Destination Stops}
	\label{fig:request_stop_type_as_destination}
\end{subfigure}
\caption{Trips over Time Period (all zones combined; West Atlanta; Belvedere).}
\label{fig:request_stop}
\end{figure}

\noindent This section explores whether riders indeed used \Reach{} to bridge the first/last mile of multimodal trips.
To help identify such trips, Figure~\ref{fig:request_stop} categorizes requests based on origin and destination, pilot zone, stop type, and time of day.
One notable pattern is the high number of connections to rail in West Atlanta.
For instance, during the morning peak, over 70\% of the riders request \Reach{} trips that end at rail stations.
In the evening peak, the situation is reversed, and about 60\% of riders request \Reach{} trips that start from rail stations.
These patterns suggest that the first/last mile problem is particularly pronounced in West Atlanta, and that ODMTS could offer a practical solution. That said, these observations stem from a pilot operating at limited scale, and accessible rail stations nearby naturally encourage multimodal behavior. Whether similar patterns would emerge in other areas warrants further investigation.
In Belvedere, the share of rail connections was lower, but still exceeded 30\% during peak hours.
This difference likely reflects the distinct characteristics of the two areas: West Atlanta is primarily low-income residential, while Belvedere is a mixed-use neighborhood with more points of interest within the service area.

Overall, the results suggest that \Reach{} improved access to the transit system for riders in under-served areas.
Even when rail stations and bus stops are within walking distance, shuttles may be preferred due to factors such as pedestrian safety, mobility limitations, carrying groceries, or harsh weather conditions.
Section~\ref{sec:mode_switching} confirms that some riders indeed switched from active modes like walking, e-scooters, and bikes.
Taken together, these findings highlight a fundamental distinction between \Reach{} and a generic ride-hailing service: \Reach{} shuttles operated exclusively within designated geographic zones, making trips inherently short and local, and the dominant use pattern was as a first/last-mile connector to fixed-route transit rather than a substitute for point-to-point transportation, which is the market of ride-hailing systems. These two services address different needs and will likely co-habit in the future.

\subsection{Quality of Service}
\label{subsect:quality}

\begin{figure}[htbp]
\centering
\begin{subfigure}[!t]{0.9\textwidth}
	\centering
	\includegraphics[trim={0 1.4cm 0 2cm}, clip, width=\textwidth]{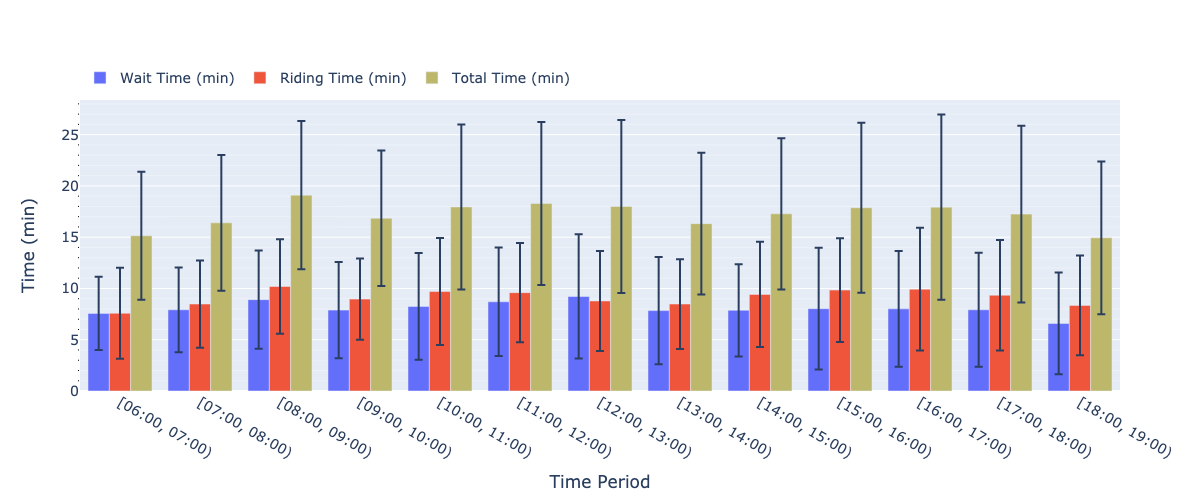}
	\caption{
		Average \Reach{} Travel Time per Time Period.
	}
	\label{fig:reach_travel_time}
\end{subfigure}
\begin{subfigure}[!t]{0.9\textwidth}
	\centering
	\includegraphics[trim={0 1cm 0 2cm}, clip, width=\textwidth]{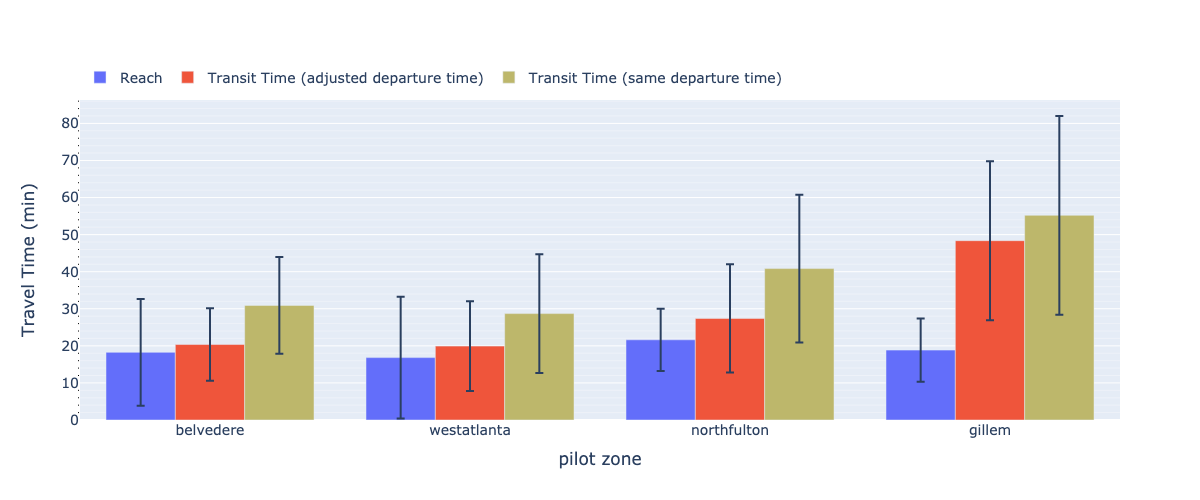}
	\caption{
		Average \Reach{} Travel Time vs. Average Estimated Transit Time.
	}
	\label{fig:reach_vs_fixed_routes}
\end{subfigure}
\caption{Key Results on Travel Time, with Standard Deviations}
\end{figure}

This section reports on a number of metrics for evaluating the quality
of \MartaReach{} services. Figure~\ref{fig:reach_travel_time} reports the average waiting time, riding time, and total travel time of \Reach{} during different service hours. The average waiting time for
\MartaReach{} was approximately 8 minutes throughout the day, while the
average riding time was less than 10 minutes. Interestingly, the travel
time remained consistent irrespective of peak hours,
despite slightly longer trips observed between 8:00 to 9:00 a.m.\ and
between 4:00 to 5:00 p.m. This is because \Reach{} primarily served the
first and last mile within local regions, which are less affected by
traffic congestion during peak hours.
The same characteristic was observed in full-scale ODMTS simulations \citep{auad2021resiliency}.

\subsubsection{Comparison with Fixed Routes}

Figure~\ref{fig:reach_vs_fixed_routes} compares the average travel time between \Reach{} and fixed-route buses in all four pilot zones. The \emph{Travel Time} for \Reach{} is computed as the duration between request time at trip origin and alight time at trip destination. The travel time for fixed routes is estimated for the same origin-destination pair, assuming that the rider will walk for their first and last legs. For \emph{Same Departure Time}, the departure time is set to match the request time of the corresponding \Reach{} trip. \emph{Adjusted Departure Time} provides an optimal fixed-route travel plan by synchronizing the rider departure time to the bus schedule. This option thus assumes that the rider has the flexibility and certainty to perfectly time their departure.

The results in Figure~\ref{fig:reach_vs_fixed_routes} show that \Reach{} trips were, on average, faster than taking the same trip with the existing transit system.
Even compared to the setting in which riders can align their departure time to the fixed transit system, \Reach{} delivered a better travel time for 60\%-64\% of the trips over the different zones. This number increases to 84\%-85\% compared to the fixed-route system with the same departure time. Also note that the reported times for \Reach{} are actual times, not estimates.
The benefits of \Reach{} were particularly pronounced in the Gillem pilot zone, because the existing fixed routes
are nearly inaccessible for most of the industrial facilities.  Note also that the differences between \textit{Adjusted Departure Time} and \textit{Same Departure Time} are significant: traveling with fixed routes requires careful synchronization, while \Reach{} provided a more flexible service to the riders than the current transit system.
The frequent rider example from Section~\ref{subsubsect:frequent_riders} suggests that it was a service dependable enough for daily use.

The short travel times observed in the pilot are promising, though it is important to note that, with many direct trips and limited ride sharing, travel times were recorded under conditions more favorable than those of a full-scale ODMTS deployment. At larger scales, higher demand would increase ride sharing, introduce detours, and may extend travel times; the 2026 \MartaReach{} deployment provides a real-world context in which future researchers can examine.
\Reach{} mostly provided direct trips, but higher demand could increase ride sharing and extend travel times.
Simulation studies suggest that even at scale, limited ride sharing may be sufficient (see Section~\ref{subsect:ridership}), helping to keep travel times low, though this remains to be confirmed in practice.
Greater demand may also reduce waiting times, as more vehicles would be spread across the area and less idling is needed.
The idling is significant, as \Reach{} vehicles spent a lot of time driving back and forth to the few designated idle locations and frequently became unresponsive there (see Appendix~\ref{sec:fleet_and_driver}).
Optimizing these idle locations, currently selected by MARTA, could improve waiting times even at smaller scales.
Operational constraints also offer room for improvement.
For safety reasons, drivers were not allowed to receive new assignments while driving, limiting responsiveness.
Enhancing the driver-app interface may help address this.
\Reach{} maintained an average waiting time of about 8 minutes, while simulations suggest that 3–7 minutes may be achievable at scale \citep{auad2021resiliency}, though these projections assume operating conditions beyond what the pilot was able to test, and should be interpreted as an informed expectation rather than a confirmed outcome.

\subsection{Analysis on Cancellations}
\label{sect:cancel}

Section~\ref{subsect:ridership} reported on a high number of cancellations, many of which were soon followed by an actual trip.
Table~\ref{tab:cancel_request} dives into this behavior in more detail for different time thresholds $\theta$ in four categories:

\begin{enumerate}
	\item \textit{Exact Return:} the rider made another request with the same origin-destination pair within $\theta$ minutes from the cancellation time and the request was served;

	\item \textit{Other Return:} the rider  made another request with a different origin-destination pair within $\theta$ minutes from the cancellation time and the request was served;

	\item \textit{Repeated Cancellations:} After canceling the first request and before getting a ride within $\theta$ minutes, the rider kept submitting requests and canceling them.

	\item \textit{No Return:} The rider cancelled the request and did not take a ride within $\theta$ minutes.
\end{enumerate}

Table~\ref{tab:cancel_request} shows that approximately 30\% of cancellations were followed by an actual
ride within one hour.
A possible explanation is that these riders had some flexibility in their schedule, and viewed \Reach{} as their preferred mode of transportation.
Secondly, it was observed that riders may slightly change their origins and destinations to try to match a vehicle.
This is highlighted by the number of cancellations classified as \emph{other return}.
These behaviors were facilitated by the high density of \Reach{} stops, with many alternative pick-up and drop-off options nearby.
Some riders were found to cancel repeatedly, presumably to game the system and attempt to minimize wait time by getting assigned to a closer vehicle.
This exploits the shuttle dispatching system, which is not based solely on proximity, but solves a real-time optimization problem based on vehicle availability, current assignments, and proximity to determine which vehicle should serve which request \citep{riley2019column}.

\Reach{} did not impose penalties or charge cancellation fees, making it easy for users to cancel at any time without consequences. Whether such system-abusing behavior becomes easier to detect or manage at a larger scale, remains an open question.Future implementations would benefit from a mechanism that discourages frequent cancellations, especially repeated cancellations intended to game the system.
It may also be valuable to study the motivations behind rider cancellations, particularly when users do not return to the service within $\theta$ minutes.
Possible reasons include:
(i) riders returned after $\theta$ minutes,
(ii) the waiting time was unacceptable,
(iii) riders were a no-show, which is recorded in the data as a cancellation,
(iv) users were testing the rider app without intending to take a ride.

\begin{table}[!t]
	\centering
    \resizebox{0.9\textwidth}{!}{
	\begin{tabular}{ c c c c c | c}
		\toprule
		$\theta$ (min) & Exact Return & Other Return & Repeated Cancellations & No Return  &  Total \\ \midrule

		15  & 461 (17.93\%) & 254 ~(9.87\%)~ & 200 ~(7.78\%) & 1657 (64.42\%) & \multirow{3}{*}{2,572 (100\%)} \\

		30  & 481 (18.70\%) & 280 (10.89\%) & 249 ~(9.68\%) & 1562 (60.73\%) \\
		60  & 488 (18.97\%) & 306 (11.90\%) & 275 (10.69\%) & 1503 (58.44\%)\\ \bottomrule
	\end{tabular}
    }
	\caption{Classification of Canceled Requests.}
	\label{tab:cancel_request}
\end{table}

Most cancellations (78.93\%) occurred while riders were waiting for their vehicles.
The remaining cancellations (21.07\%), were categorized as no-shows.
Among the 2,038 cancellations during the riders' waiting period, 564, 313, and 206 occurred after five, eight, and ten minutes of waiting time, respectively.
Considering that the average waiting time for served requests throughout the entire pilot was around eight minutes, it is possible that these aforementioned cancellations occurred due to riders observing that the assigned vehicles were not approaching them.
This raises questions regarding the performance of drivers and fleets during the pilot, which is discussed in Appendix~\ref{subsect:driver_behavior}.

\subsection{Economic Analysis}
\label{subsect:operating_costs}

This section provides an economic analysis that is divided in two
	parts. The first part dives into the relationship between \Reach{}
	riders and their income. The second part takes a closer look at the
	costs of the pilot under multiple scenarios and an economic lens.

\subsubsection{Rider Income and Trips: Inference from Data}

The income analysis leverages two sources of data: the long-term \Reach{} evaluation survey and the American Community Survey in the pilot year \citep{census2022, Census2022Counties}.
The long-term evaluation survey included the following question:
	\textit{“Which of the following categories best represents your
		total annual household income before taxes?”} A total of 232
	records are available for this question: 86 riders answered it, 26
	reported that they preferred not to answer it, and 120 did not answer it.
    Up to 83\% of those who answered the question live in a household
	whose annual income is strictly less than the mean annual household
	income (\$114,862) of Atlanta Metropolitan Area \citep{census2022}.

	Since half of the \Reach{} riders did not provide their income levels,
	it is important to estimate their \textit{individual} income level
	indirectly through the trip data.  The resulting estimations can then
	be compared to the \textit{mean income per capita} of the Atlanta
	Metropolitan Area in 2022, which is \$43,561 \citep{census2022}. To
	produce the estimated annual income of a \Reach{} trip, the origin and
	destination census tracts of each trip were computed and the mean
	income per capita of each individual census tract was retrieved from
	the 2022 American Community Survey \citep{Census2022Counties}. For
	those trips whose origin and destination were both virtual \Reach{}
	stops or both existing MARTA rail or bus stops, the estimated income
	is the average of the census tracts of the origin and the destination.
	For those trips where the origin or destination is a virtual \Reach{}
	stop and the other is an existing MARTA bus or rail stop, the trip can
	be assumed to be multi-modal and only the mean income per capita of
	the census tract of the \Reach{} virtual stop is considered for the
	estimation.

	The results are presented in Table \ref{tab: Estimated
		annual income}. The income range from \$40,000 to \$49,999,
	presented in blue, represents the mean income per capita in Atlanta
	Metropolitan area.
	Observe that up to 67\% of the \Reach{} trips
	were performed by a user whose estimated income is strictly less
	than the mean income per capita in Atlanta Metro. This percentage is
	100\% in West Atlanta, 10\% in Belvedere, 100\% in Gillem, and 36\%
	in North Fulton.

\begin{table*}[!t]
	\caption{Estimated annual income of trips' main user and empirical accumulated distribution.}
	\resizebox{\linewidth}{!}{%
		\centering
		\begin{tabular}{l r r r r r r r r r r}
			\toprule
			\multirow{2}*{\makecell{\textbf{Estimated annual income}\\ \textbf{range of trips' main user}}} & \multicolumn{5}{c}{\textbf{Trip counts}} & \multicolumn{5}{c}{\textbf{Accumulated percentage (\%)}} \\
			\cmidrule(lr){2-6} \cmidrule(lr){7-11}
			& \textbf{All zones} & \textbf{West Atlanta} & \textbf{Belvedere} & \textbf{Gillem} & \textbf{North Fulton} & \textbf{All zones} & \textbf{West Atlanta} & \textbf{Belvedere} & \textbf{Gillem} & \textbf{North Fulton} \\
			\midrule
			Under \$20,000 & 902 & 892 & 0 & 9 & 1 & 11.87 & 20.09 & 0 & 2.85 & 0.51 \\
			\$20,000 to \$24,999 & 1,531 & 1,205 & 0 & 304 & 22 & 32.03 & 47.23 & 0 & 99.05 & 11.73 \\
			\$25,000 to \$29,999 & 1,304 & 1,288 & 8 & 3 & 5 & 49.2 & 76.24 & 0.3 & 100 & 14.29 \\
			\$30,000 to \$34,999 & 1,095 & 918 & 148 & 0 & 29 & 63.61 & 96.91 & 5.9 & 100 & 29.08 \\
			\$35,000 to \$39,999 & 271 & 137 & 119 & 0 & 15 & \textbf{\underline{67.18}} & \textbf{\underline{100}} & \textbf{\underline{10.4}} & \textbf{\underline{100}} & \textbf{\underline{36.73}} \\
			{\color{blue}\$40,000 to \$49,999*} & {\color{blue}532} & {\color{blue}0} & {\color{blue}493} & {\color{blue}0} & {\color{blue}39} & {\color{blue}74.18} & {\color{blue}100} & {\color{blue}29.05} & {\color{blue}100} & {\color{blue}56.63} \\
			\$50,000 to \$59,999 & 451 & 0 & 440 & 0 & 11 & 80.12 & 100 & 45.69 & 100 & 62.24 \\
			\$60,000 to \$69,999 & 1,166 & 0 & 1,158 & 0 & 8 & 95.47 & 100 & 89.49 & 100 & 66.33 \\
			\$70,000 to \$79,999 & 296 & 0 & 278 & 0 & 18 & 99.37 & 100 & 100 & 100 & 75.51 \\
			\$80,000 to \$89,999 & 47 & 0 & 0 & 0 & 47 & 99.99 & 100 & 100 & 100 & 99.49 \\
			\$90,000 or more & 1 & 0 & 0 & 0 & 1 & 100 & 100 & 100 & 100 & 100\\
			\midrule
			Totals & 	7,596 & 	4,440 & 	2,644 & 	316 & 	196 \\
			\cmidrule(lr){1-6}
			Population & 89,123	 & 	21,031 & 	24,816 & 	20,264 & 	23,012 \\
			\cmidrule(lr){1-6}
			\multicolumn{11}{l}{* Contains the mean income per capita in Atlanta Metropolitan Area: \$43,561 \citep{census2022}}
		\end{tabular}
	}
	\label{tab: Estimated annual income}
\end{table*}

\begin{figure}[htbp]
	\begin{subfigure}[ht]{0.5\textwidth}
		\centering
		\includegraphics[width=0.9\textwidth]{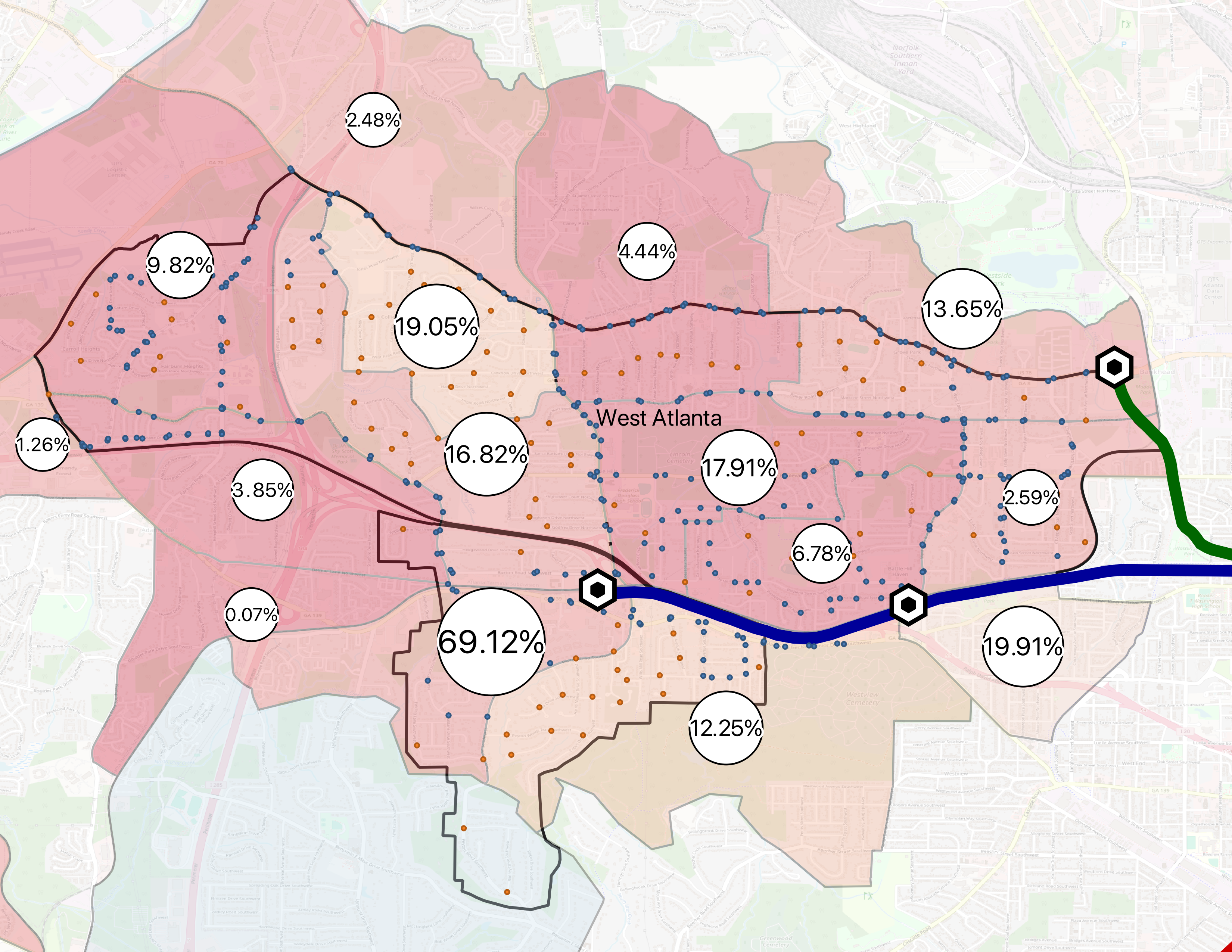}
		\caption{West Atlanta}
		\label{fig:Utilization West Atlanta}
		\vspace{0.5\baselineskip}
	\end{subfigure}
	\begin{subfigure}[ht]{0.5\textwidth}
		\centering
		\includegraphics[width=0.9\textwidth]{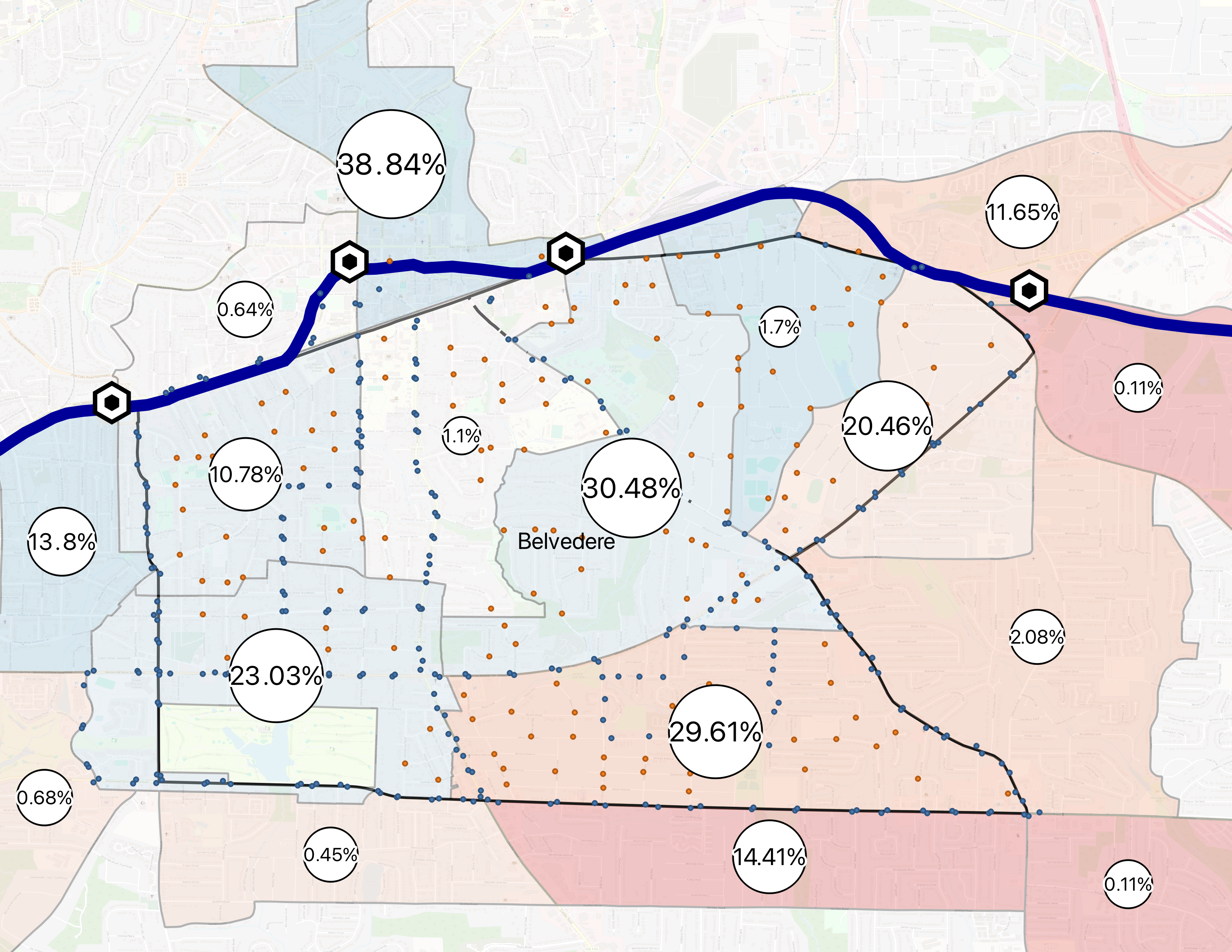}
		\caption{Belvedere}
		\label{fig:Utilization Belvedere}
	\end{subfigure}
	\begin{subfigure}[ht]{0.5\textwidth}
		\centering
		\includegraphics[width=0.9\textwidth]{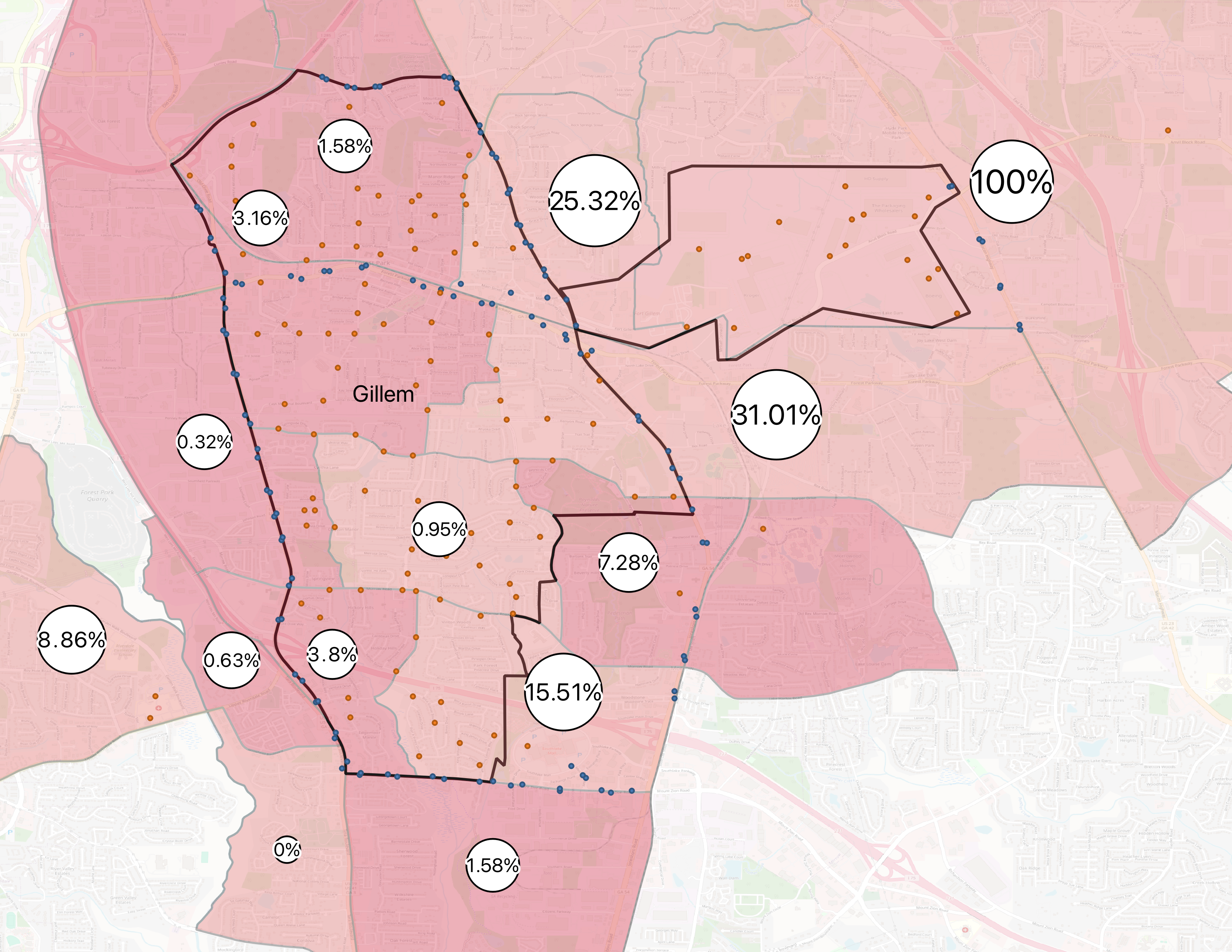}
		\caption{Gillem}
		\label{fig:Utilization Gillem}
		\vspace{0.5\baselineskip}
	\end{subfigure}
	\begin{subfigure}[ht]{0.5\textwidth}
		\centering
		\includegraphics[width=0.9\textwidth]{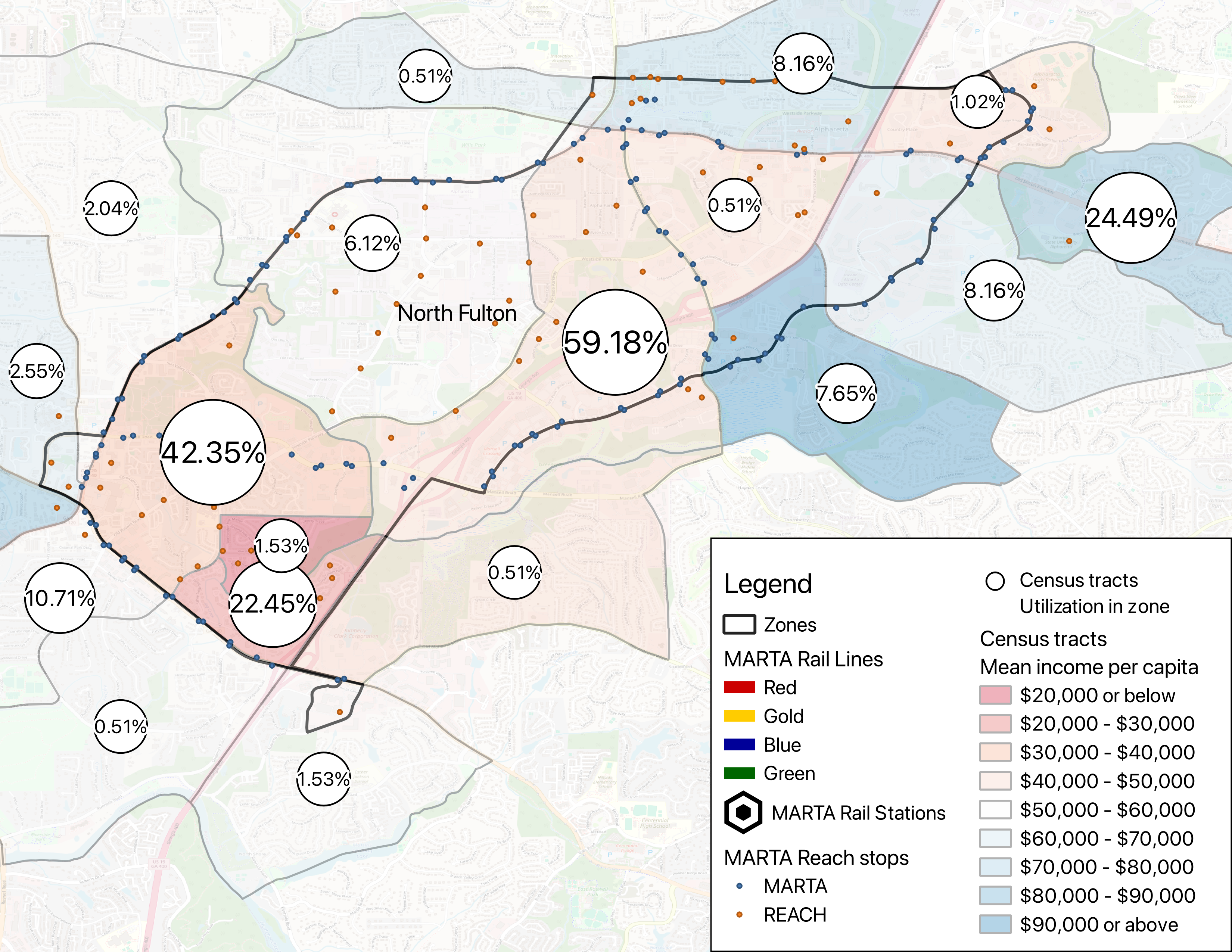}
		\caption{North Fulton}
		\label{fig:Utilization North Fulton}
	\end{subfigure}
	\caption{Utilization of the census tracts in each zone with their mean income per capita.}
	\label{fig:census_tract_utilization}
\end{figure}

The next results consider the distribution of trips among census tracts: they are computed as follows:
\begin{equation}
    \resizebox{0.5\textwidth}{!}{
	$
		\text{census tract utilization} = \frac{\text{number of trips that started / ended in the census tract}}{\text{number of trips in the zone}}
	$
    }
\end{equation}
	The results are plotted in Figure \ref{fig:census_tract_utilization},
	along with the mean income per capita of each census tract. The two
	zones with the largest ridership are West Atlanta and Belvedere, the
	two pilot zones with direct access to MARTA rail stations. This
	explains why the majority of the \Reach{} trips happen in these two
	zones: Figure \ref{fig:census_tract_utilization} highlights that the
	census tracts around rail station have higher utilization. Moreover,
	although the West Atlanta population (21,031) is lower than the
	population in Belvedere (24,816), ridership in the former (4,440) is
	much larger than in the latter (2,644). Table \ref{tab: Estimated
		annual income} and Figure \ref{fig:census_tract_utilization} give a
	possible explanation: trips in West Atlanta are performed by riders
	with lower income than those from Belvedere.

    In summary, the results in this section are interesting for two
	reasons. First, \Reach{} seems to have benefitted low-income populations,
	providing a service that was desirable for their
	mobility needs. Second, the service remained attractive for other
	population segments as exhibited in Belvedere, which represent higher income levels.
	Indeed, a significant number of riders in
	Belvedere have an estimated average individual income between
	\$60,000 and \$70,000.

\subsubsection{Cost structure}

It is also interesting to quantify the cost of the \Reach{} service to
inform transit agencies about the tradeoffs between quality of service
and costs. These tradeoffs were computed using the simulator from
\citet{auad2021resiliency}: the simulator was run for a full day of
trips, computing costs and waiting times for various scenarios based on the observations of the pilot. The cost per vehicle hour
was multiplied by the number of hours of service and the number of
vehicles, before being divided by the number of riders served to
derive the cost per rider per hour for the MARTA \Reach{}
service.

Table~\ref{tab:cost-per-rider-ridership-westatlanta} presents
the sensitivity analysis for different costs per vehicle hour and for
different fleet sizes and ridership levels for the West Atlanta zone.
Given the continuously growing ridership observed in the pilot (see Section~\ref{subsect:ridership}), two potential growth scenarios are included in the analysis.
The pilot used 6 shuttles in this zone, but results are also
presented for 5 and 7 shuttles. The waiting times are presented at the
top of the table for each fleet size and demand scenario. For the
demand scenarios, the simulator uses requests from August 31, 2022,
the pilot day with the highest ridership, and from recent days preceding
August 31, 2022, until a desired ridership level is reached in the
case of growth scenarios. Trips that involve a rider from August 31,
2022 on a previous day are only sampled for increased ridership if the
time window of the trip on the earlier day does not overlap with the
time window of that rider's trip on August 31, 2022.

\begin{table}[!t]

	\centering
    \caption{Cost (\$) per rider in West Atlanta Across Ridership Levels and Fleet Sizes}
	\resizebox{\columnwidth}{!}{
		\begin{tabular}{cccclccclccc}
            \toprule
			& \multicolumn{11}{c}{Cost per Rider (\$)}                                                                   \\ \cmidrule{2-12}
			\multicolumn{1}{c}{\multirow{3}{*}{\begin{tabular}[c]{@{}c@{}}Cost per\\ Vehicle/hour\\ (\$/hr)\end{tabular}}} & \multicolumn{3}{c}{Base Ridership}       &  & \multicolumn{3}{c}{2x Ridership}         &  & \multicolumn{3}{c}{3x Ridership}         \\ \cmidrule{2-4} \cmidrule{6-8} \cmidrule{10-12}
			\multicolumn{1}{c}{}                                                                                           & \multicolumn{3}{c}{Fleet Size (Wait Time)} &  & \multicolumn{3}{c}{Fleet Size (Wait Time)} &  & \multicolumn{3}{c}{Fleet Size (Wait Time)} \\ \cmidrule{2-4} \cmidrule{6-8} \cmidrule{10-12}
			\multicolumn{1}{c}{}                                                                                           & 5 (7.75 min)        & 6 (7.26 min)       & 7 (7.36 min)        &  & 5 (9.81 min)         & 6 (8.50 min)       & 7 (7.80 min)      &  & 5 (17.35 min)        & 6 (10.85 min)       & 7 (9.04 min)      \\ \midrule
			20                                                                                                             & 14.61    & 17.53    & 20.45    &  & 7.14     & 8.57    & 10.00   &  & 4.81     & 5.78    & 6.74   \\
			25                                                                                                             & 18.26    & 21.91    & 25.56    &  & 8.93     & 10.71    & 12.50   &  & 6.02     & 7.22    & 8.43   \\
			30                                                                                                             & 21.91    & 26.29    & 30.67    &  & 10.71     & 12.86    & 15.00   &  & 7.22     & 8.67    & 10.11   \\
			35                                                                                                             & 25.56    & 30.67    & 35.79    &  & 12.50     & 15.00    & 17.50   &  & 8.43     & 10.11    & 11.80   \\
			40                                                                                                             & 29.21    & 35.06    & 40.90    &  & 14.29     & 17.14    & 20.00   &  & 9.63     & 11.56    & 13.48   \\
			45                                                                                                             & 32.87    & 39.44    & 46.01    &  & 16.07     & 19.29    & 22.50   &  & 10.83     & 13.00    & 15.17   \\
			\midrule
			50                                                                                                             & 36.52    & 43.82    & 51.12    &  & 17.86     & 21.43    & 25.00   &  & 12.04     & 14.44    & 16.85   \\
			55                                                                                                             & 40.17    & 48.20    & 56.24    &  & 19.64     & 23.57    & 27.50   &  & 13.24     & 15.89    & 18.54   \\
			60                                                                                                             & 43.82    & 52.58    & 61.35    &  & 21.43     & 25.71    & 30.00   &  & 14.44     & 17.33    & 20.22   \\
			65                                                                                                             & 47.47    & 56.97    & 66.46    &  & 23.21     & 27.86    & 32.50   &  & 15.65     & 18.78    & 21.91   \\
			70                                                                                                             & 51.12    & 61.35    & 71.57    &  & 25.00     & 30.00    & 35.00   &  & 16.85     & 20.22    & 23.59   \\
			\midrule
			75                                                                                                             & 54.78    & 65.73    & 76.69    &  & 26.79     & 32.14    & 37.50   &  & 18.06     & 21.67    & 25.28   \\
			80                                                                                                             & 58.43    & 70.11    & 81.80    &  & 28.57     & 34.29    & 40.00   &  & 19.26     & 23.11    & 26.96   \\
			85                                                                                                             & 62.08    & 74.49    & 86.91    &  & 30.36     & 36.43    & 42.50   &  & 20.46     & 24.56    & 28.65   \\
			90                                                                                                             & 65.73    & 78.88    & 92.02    &  & 32.14     & 38.57    & 45.00   &  & 21.67     & 26.00    & 30.33   \\
			95                                                                                                             & 69.38    & 83.26    & 97.13    &  & 33.93     & 40.71    & 47.50   &  & 22.87     & 27.44    & 32.02   \\
			100                                                                                                            & 73.03    & 87.64    & 102.25   &  & 35.71     & 42.86    & 50.00   &  & 24.07     & 28.89    & 33.70   \\ \bottomrule
	\end{tabular}}
	\label{tab:cost-per-rider-ridership-westatlanta}
\end{table}

The economic analysis in
	Table~\ref{tab:cost-per-rider-ridership-westatlanta} shows that the
	cost per rider quickly goes down as demand increases.  For the
shuttles used in the \MartaReach{} pilot, the cost per vehicle hour is
expected to be between \$50 to \$70 an hour.  This results in a
	high cost of between \$36.52 (5 vehicles at \$50/hour) and \$71.57
	(7 vehicles at \$70/hour) per rider at the ridership level observed
	during the pilot.  However, at twice or three times the pilot ridership, \Reach{} is simulated to
	operate at a substantially lower cost of \$17.86--\$35.00 per rider
	or \$12.04--\$23.59 per rider, respectively. A potentially more sustainable path forward is to use smaller vehicles, like sedans. This was already discussed in Section~\ref{subsect:ridership} for capacity reasons, as ride sharing with more than three people was rarely observed.

	Ride-hailing services such as Uber have a cost of approximately
	\$28.21 per hour when the average labor (\$19.36) and the vehicle
operating cost (\$8.85) are combined
\citep{Helling2020-Ridesters2020Independent,
	helling_stephen_hempler_paul_williams_savage_tom_2023}. Operating
\Reach{} with similar vehicles would result in a cost in the ranges
\$18.26--\$30.67, \$8.93--\$15.00, and \$6.02--\$10.11 per rider for the pilot
ridership, twice the ridership, and three times the ridership
respectively.  To put this into context, public transit is
	typically subsidized and fare revenues only pay for part of the
	operating expenses.  In the case of MARTA, fares only amounted to
	12.2\% of the operating expenses in 2022
	\citep{FTA2022-NationalTransitDatabase}.  At a cost of
	\$6.02--\$10.11, this corresponds to a fare collection of
	\$0.74--\$1.23 per trip, much less than the current ticket price.
	This suggests that ODMTS may be economically viable, even with a
	relatively small ridership (3x the pilot ridership) and when serving
	a difficult, low-income area that currently has limited bus coverage
	(West Atlanta).  If the ridership continues increasing,
	\citet{auad2021resiliency} estimate that the cost may be brought all
	the way down to \$2.27 per rider when the system is deployed at city
	scale.

    Table~\ref{tab:cost-per-rider-ridership-westatlanta} is intended to show how the agency's cost per rider improves with scale, not to serve as a direct cost comparison with TNC services. Two key differences are worth noting here. First, \MartaReach{} operated large paratransit vehicles rather than sedans as the TNCs, which significantly inflated the cost per vehicle hour, as discussed earlier. Second, while TNCs may appear to have reasonably low operating costs, those costs are ultimately recovered through market-rate fares charged to riders; also, expenses such as vehicle maintenance are covered by individual drivers rather than the platform itself. \Reach{} and ODMTS charge a low and flat transit fare, with the agency covering all cost themselves through public subsidy. A lower cost per rider in Table~\ref{tab:cost-per-rider-ridership-westatlanta} does not imply that the transit agency will start to generate profit; this remains a public service, and any savings benefit the agency and the riders it serves. Additionally, these projections are based on simulation and assume operating conditions at a scale not yet validated in practice; confirming them requires larger deployments and continued empirical investigation.

\section{Conclusions}
\label{sect:conclusion}

This paper presented the design, implementation, and evaluation of \MartaReach{}, a six-month on-demand mobility pilot conducted in Atlanta, GA in 2022.
\Reach{} aimed to address key open questions about the practical viability of On-Demand Multimodal Transit Systems (ODMTS).
ODMTS adopt a transit-centric model, integrating fixed-route transit with on-demand shuttles to provide integrated, door-to-door multimodal service.
Although simulation studies have demonstrated the theoretical potential of ODMTS, the system has not yet been tested in practice at scale.
\Reach{} offered on-demand services across four zones with distinct characteristics, connecting each zone to the existing transit network and offering an integrated flat fare.
This allowed for testing key system components in real-world conditions and revealed practical insights toward the realization of ODMTS.

Since its launch, \Reach{} experienced steady ridership growth throughout the pilot, continuing until the final day. The service was widely appreciated, with 94\% of riders indicating to be satisfied or very satisfied.
Most \Reach{} users were already transit users, but 35\% of trips would not have been transit trips otherwise.
A substantial number of mode shifts were observed away from cars, taxis, and ride-hailing services.
Many trips were multimodal, especially in West Atlanta, where over 70\% of morning riders connected to the rail.
This provides supporting evidence that ODMTS can attract new riders and that travelers are open to making multimodal trips.
That said, it is worth emphasizing that these observations reflect a pilot operating with limited ride-sharing and a concentrated early adopter base, and their generalizability to a full-scale should be investigated in the future.

\Reach{} trips were on average faster than taking the same trip with the existing transit system.
These short travel times are promising for full-scale ODMTS deployments, and the paper discusses several factors to keep in mind as the system is scaled.
\Reach{} maintained an average waiting time of about 8 minutes, while simulations suggest this may go down to 3-7 minutes at full scale.
The pilot also revealed several practical challenges that serve as useful lessons for future deployments.
This includes riders who repeatedly cancel trips to reduce their waiting time, challenges related to driver and fleet performance, and vehicles that were unnecessarily large.
An economic analysis showed that approximately 60\% of riders earned below the metro area's per capita income, suggesting strong appeal in lower-income neighborhoods such as West Atlanta.
However, the service also attracted higher-income users, particularly in the Belvedere zone, indicating broad demographic appeal.
Additionally, ridership patterns suggest potential for economic viability, especially when deploying smaller vehicles.

Overall, \Reach{} has provided early evidence suggesting that ODMTS may address critical mobility gaps and warrants further exploration.
The pilot also offered actionable insights into operating new mobility services in urban environments.
A key contribution is that the operational challenges encountered, including driver behavior, fleet sizing, and cancellation management, are lessons that simulation studies alone cannot easily provide and that open future directions for research on real-world ODMTS deployment.
Similar pilots can be launched in other cities to assess local needs and conditions, and to work toward the successful deployment of a city-scale ODMTS.

In summary, this pilot has yielded substantial evidence and encouraging mobility outcomes; however, certain aspects can only be partially informed by this pilot and require confirmation through larger scale deployments. In particular, whether rider adoption and the observed mode shift away from private vehicles persist as the service scales, whether service quality is maintained as demand and ride sharing increase, and whether operational challenges such as driver management, cancellations, and rider gaming intensify at scale remain questions that this pilot cannot fully answer.
Perhaps the most meaningful measure of the pilot success and benefits is the recent decision by MARTA to adopt the concept widely. On March 7, 2026, MARTA officially launched \MartaReach{} as a long-term, system-wide on-demand transit service across 12 zones \citep{MARTA-Reach-website}. The 2026 service retained the same brand name, the same integrated flat fare (matching MARTA's standard bus and rail fare with free transfers), and the same organizational model under MARTA's operational control.
This transition from a six-month academic collaboration to a long-term service provides external validation of the practical value of the pilot.
The practical insights documented in this paper, spanning ridership behavior, operational challenges, and the economic path to viability, were not only academic outputs; they also contributed to a business decision by a large transit agency, that now provides the additional mobility services reported in this paper.

\section*{Acknowledgments}
This research was partly supported by NSF CIVIC grant NSF-2133342
(deployment), NSF Leap-HI 1854684 (supporting infrastructure), and NSF AI Institute for Advances in Optimization (Awards 2112533 and 2620485). The authors
would like to thank MARTA for a successful collaboration, in
particular Robert Goodwin for the initial concept and Anthony Thomas,
Jonathan Weaver, Erick Knowles, Chris Wyczalkowski, David Emory, and the rest of the MARTA Reach team at MARTA, for
the daily operations and planning activities. The authors would also
like to thank Debra Lam and her team at Partnership for Innovation (PIN) for facilitating these
research and partnership opportunities.

\DeclareRobustCommand{\VAN}[3]{#3}   

\bibliographystyle{trc}
\bibliography{reference}

\clearpage
\appendix

\section{ODMTS Technical Research Summary}
\label{sect:odmts_additional}

The initial theoretical exploration of designing an ODMTS network (comprising both new high-frequency fixed routes and on-demand shuttles), utilizing Benders Decomposition, was presented by \citet{maheo2019benders}. Continuing along this direction, \citet{DalmeijerVanHentenryck2020-TransferExpandedGraphs} extended the previous research by introducing additional realistic considerations to the problem, such as incorporating transfer limits and preserved urban rails. \citet{auad2022ridesharing} include shuttle ride-sharing and fleet-sizing considerations in this network design problem. In contrast to the works previously mentioned, which focused on network designs with fixed transit demand, four studies have extensively investigated the incorporation of latent demand to ODMTS. The original problem was proposed by \citet{basciftci2020bilevel} as a bi-level Mixed-Integer Programming, then it was further investigated and was addressed using a Benders Decomposition-based algorithm \citep{basciftci2022capturing}. Two subsequent studies addressed the same problem using model reformulations \citep{guan2024path} and heuristic algorithms \citep{guan2025heuristic}; both approaches are significantly more computationally efficient. Building upon some of the aforementioned work, \citet{lu2024impact} investigated the effects of dedicated bus lanes on the ridership and transit times of an ODMTS. Moving from design to operations, \citet{riley2019column} investigated the real-time dispatching of on-demand shuttles using a column-generation-based algorithm. Subsequent research incorporates machine learning for demand prediction to enhance the algorithm's performance \citep{riley2020real}. Moreover, \cite{auad2021resiliency} provided a full pipeline for the design and evaluation of ODMTS, combining advanced methods for trip chaining \citep{BarryEtAl2002-OriginDestinationEstimation}, ODMTS design, fleet sizing, and real-time shuttle dispatching. Additionally, readers are also referred to \citet{hwang2025measuring} for a recently investigation into the equity and equality impacts of ODMTS. Motivated in part by the driver behavior challenges observed in the \MartaReach{} pilot, \citet{lu2026impact} investigate the potential impact of shared autonomous vehicles in microtransit systems, using data and operational insights from \MartaReach{} to quantify the benefits of removing human driver latency from the dispatching loop.

\section{Timeline of Key Events}
\label{sec:timeline}

\begin{figure}[htbp]
	\centering
	\fbox{\includegraphics[width=\textwidth]{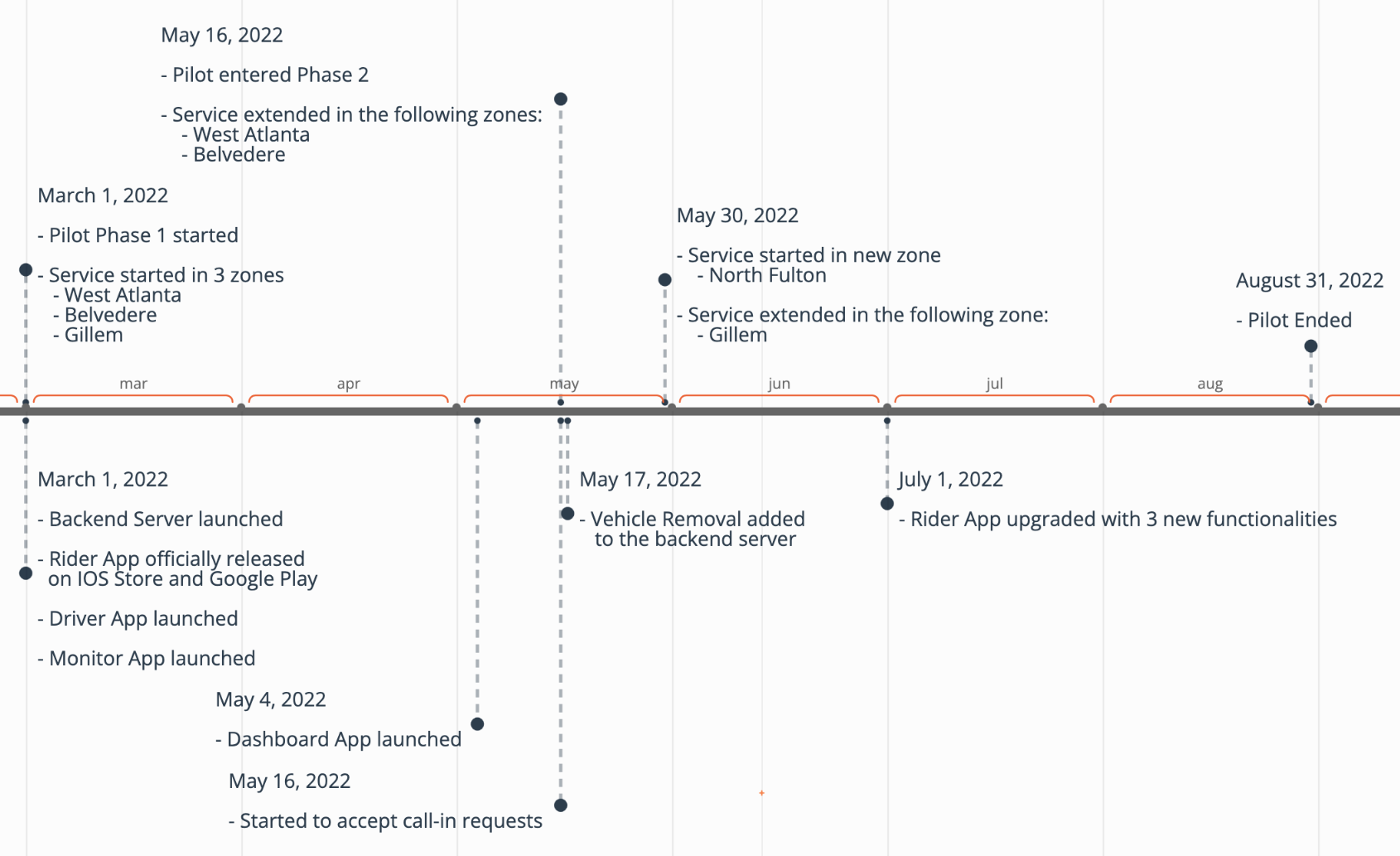}}
	\caption{Timeline for \MartaReach{} Pilot.
	}
	\label{fig:timeline}
\end{figure}

\noindent Figure~\ref{fig:timeline} summarizes the pilot timeline, with a
particular focus on the critical events for operations and technology
developments that occurred during the implementation. {\em The pilot
	was a fertile ground for rapid deployment and experimentation}: the
\Reach{} team worked at a fast pace to address challenges occurring in
the field and added new functionalities to improve the rider experience. Note that this timeline includes only events occurring during the six-month pilot period; pre-pilot and post-pilot activities such as zone selection, data analysis, and post-pilot survey collection are not included.
The pilot was performed in two phases.
Phase~1 ran from the pilot launch on March 1, 2022 to May 15, 2022 and served three zones: West Atlanta, Belvedere, and the Gillem Logistics Center.
For most of this time, the pilot was operated with a rider mobile application, a driver mobile application, and a monitor web application.
A dashboard web application was introduced near the end of the first phase.
The start of Phase~2 was marked by an extension of the three pilot zones on May 16, 2022, and continued until the end of the pilot on August 31, 2022.
Furthermore, the new North Fulton zone was added on May 30, 2022.

Phase~2 also expanded the technology infrastructure in significant ways. While most riders used the mobile application, call-in requests were introduced in Phase~2 to ensure that individuals who may not have access to a smartphone or prefer not to use it could still utilize the \MartaReach{} service. In Phase~2, the rider application was
upgraded with three new features to provide a more user-friendly
experience: it allowed riders to request frequent trips with one
click, gave riders shortcuts to frequent origins and destinations, and
enabled users to find virtual stops based on street address. The first
feature eliminated the need to manually enter the same information
repeatedly. The second feature enabled riders to easily select
frequent stops associated with their account. The third feature
introduced an address search bar to help riders locate virtual stops
near their origin/destination address. This feature made it easier for
riders to identify the best stop to use, particularly when they were
unfamiliar with the area.

Recalling Section~\ref{sect:introduction}, answering some of the remaining open questions will require observing larger-scale deployments, and the pilot's success indeed encouraged MARTA to relaunch \MartaReach{} system-wide in March 2026. However, the authors are not collaborating with MARTA on this expanded service; therefore, the specific milestones of the large-scale service fall outside the scope of this paper and are not presented in the timeline.

\section{\Reach{} Technology}
\label{sect:technology}

This section presents the four mobile/web applications used during
\Reach{}: the rider mobile application, the driver mobile application,
the monitor application, and the dashboard. All of these apps are
developed with React and React Native, popular tools for building user
interfaces (UI). React is a JavaScript library based on UI components
and React Native is a UI software framework. The section also presents
the backend which dispatches vehicles and supports the above mentioned
applications. After conducting initial testing in February 2022, all
applications (except the dashboard) were officially launched on the
first day of the pilot (March 1, 2022). The dashboard was released
on May 4, 2022 before the pilot entered Phase~2. During the pilot, in
order to proactively address feedback from both riders and dispatchers
and to overcome challenges that arose, multiple extensions were
introduced to the applications. In addition to the key additions shown
in Figure~\ref{fig:timeline}, other significant extensions and their
respective release dates are highlighted in this section.

\subsection{The Rider Application}
\label{subsect:riderapp}

\begin{figure}[htbp]
	\centering
	\includegraphics[width=\linewidth]{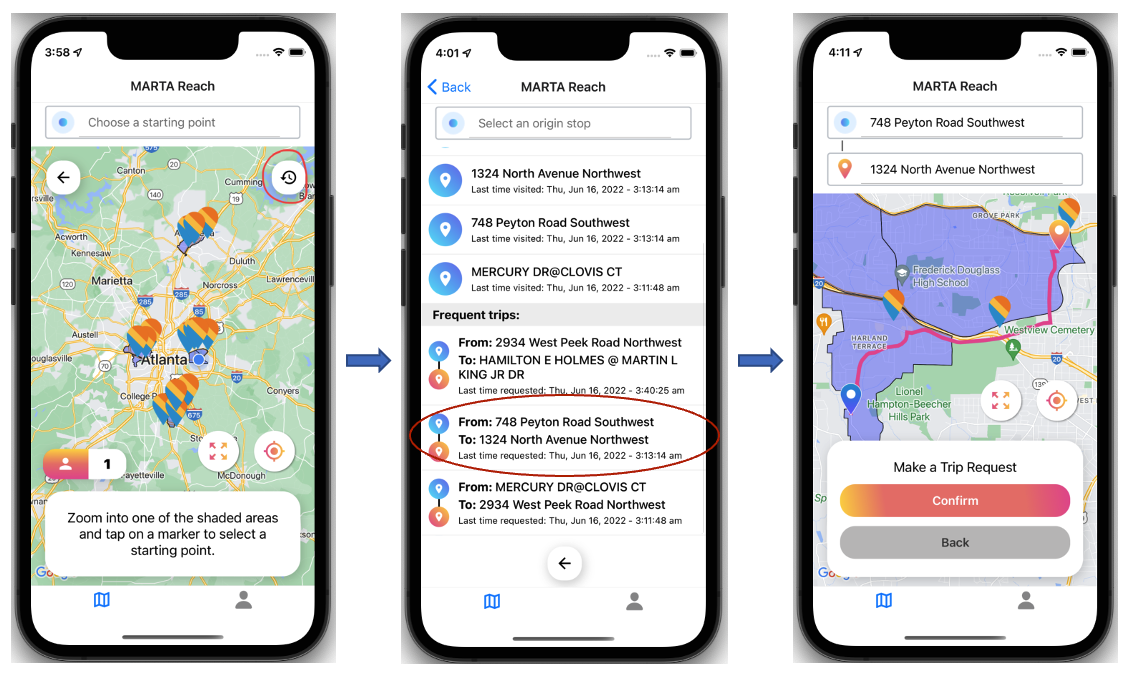}
	\caption{The Rider Application and the Selection of Frequent Trips.}
	\label{fig:ride-request}
\end{figure}

As illustrated in Figure~\ref{fig:ride-request}, the rider app is a
user-friendly mobile app that enables riders to quickly and easily
hail a ride from their current location to their desired destination
within one of the designated pilot zones. The application highlights
key transfer points where riders could connect to MARTA rail and bus
lines. It allows users to create an account, log in, and select their
virtual pickup location by either searching for it on the map or using
the search bar. Similarly, riders can enter their destination in the
same way. To address feedback from riders, more detailed instructions
and three new functionalities were added in June 2022. These
additions make it even more convenient for frequent users to book a
ride, track their driver arrival, and enjoy a seamless journey to
their destination.  After beta testing these new functionalities for a
few weeks, the rider application was officially upgraded on July 1,
2022.

\subsection{The Driver Application}
\label{subsect:driverapp}

\begin{figure}[htbp]
	\centering
	\includegraphics[width=\linewidth]{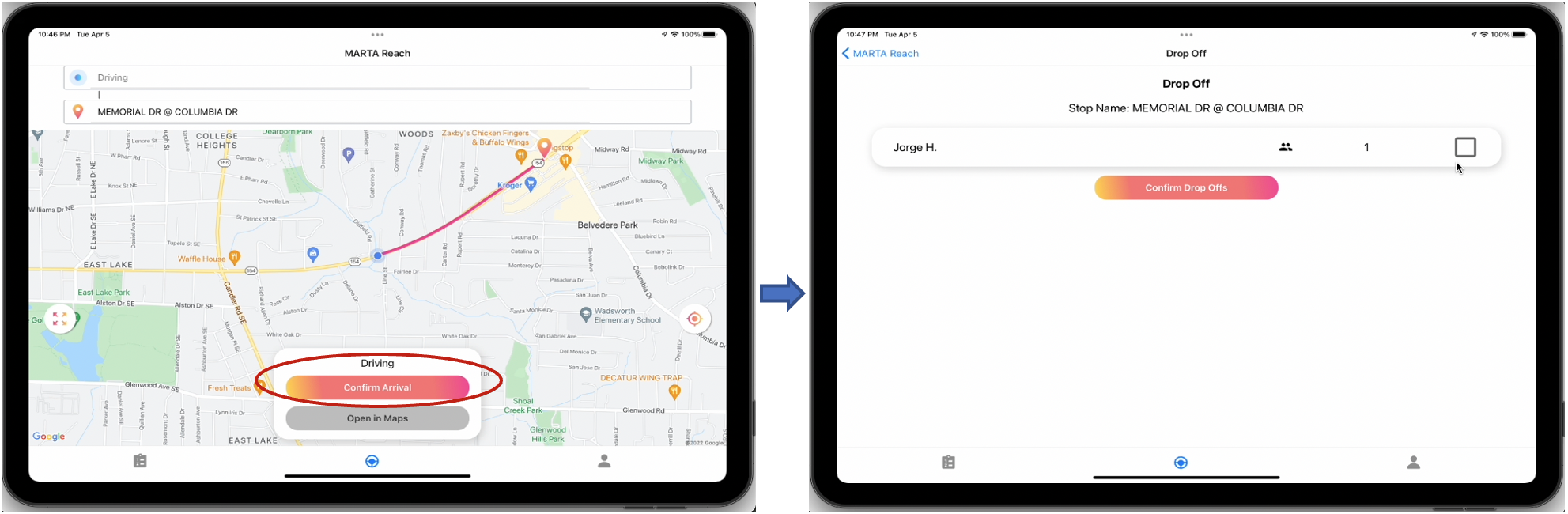}
	\caption{Driver App and Arrival Confirmation to Drop Off Riders.}
	\label{fig:driver-dropoff}
\end{figure}

The driver application manages the routes, pickups, drop-offs, and
relocations.  Once they log in, drivers wait for ride requests. From
that moment on, they will receive and serve pickup and drop-off
assignments until they request a break, their shift ends, or there are
no more requests. Drivers only see one leg at a time, i.e., the next
stop where riders will be picked up or dropped off. The application
also never communicates with them during driving, except for the map
directions. Drivers confirm pickups and drop-offs, and report no-show
passengers directly within the application. Once a driver has
completed all of their assignments, they are directed back to an idle
location to wait for another assignment.
Figure~\ref{fig:driver-dropoff} presents an example of how drivers
confirm arrival at their next stops and confirm drop-offs. The driver
application is also used to track the location of the vehicle and
provide real-time information to riders. This helps to ensure that
passengers can track their driver arrival and estimated time of
arrival (ETA) to the pickup location.

\subsection{Monitor}
\label{subsect:monitor}

\begin{figure}[htbp]
	\centering
	\begin{subfigure}[!ht]{0.9\textwidth}
		\includegraphics [width=\linewidth]{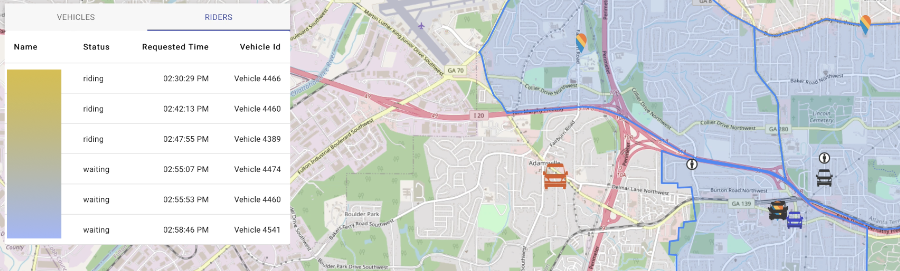}
		\caption{Monitor only shows idle stops.}
		\label{fig:monitor_1}
	\end{subfigure}
	\begin{subfigure}[!ht]{0.9\textwidth}
		\includegraphics [width=\linewidth]{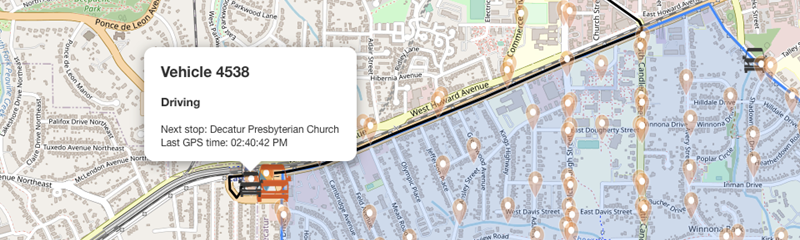}
		\caption[size=\normalsize]{Monitor shows all stops and a polyline.}
		\label{fig:monitor_2}
	\end{subfigure}
	\caption{Two Screenshots from the Monitor App, Private Information is Hidden.}
\end{figure}

The \Reach{} monitor is a web-based application that enables
dispatchers to track the shuttle fleet, view real-time key events, and
interact with the
system. Figures~\ref{fig:monitor_1}~and~\ref{fig:monitor_2} show the
fleet being monitored with real-time GPS coordinates. The vehicles
have three statuses, represented by black, blue, and red icons, which
indicate \textit{regular}, \textit{with riders}, and \textit{wrong
	location}, respectively. In particular, the \textit{regular} status
can be further divided into four subcategories---\textit{idling},
\textit{waiting for departure}, \textit{waiting for passengers}, and
\textit{driving without passengers}. The \textit{waiting for
	departure} status indicates a driver receives a request or a
rebalancing command but has not responded yet. The \textit{wrong
	location} status was added to the monitor on March 29, 2022, to
further assist the dispatchers. A vehicle is categorized as
\textit{wrong location} if a driver reported being at a stop that is
farther than 400 meters from its GPS location. In such situations, the
dispatcher would contact the driver to make adjustments. Moreover, the
dispatcher can manually sign off a vehicle using the monitor.

Riders and their assigned vehicles are displayed on the monitor once
they submit requests. Riders have two statuses in
general---\textit{waiting} and \textit{riding}, the status switches
from \textit{waiting} to \textit{riding} once the driver confirms the
pickup through the driver app. Clicking on rider icon or vehicle icon
on the monitor displays a polyline indicating the trip path. The
monitor app also presents zones and stops in the background, with zone
areas and idle stops displayed by default, and other stops being
shown by toggling check-boxes. In addition to real-time visualization,
since May 16, 2022, the monitor app includes a table that allows
dispatchers to book requests for riders who prefer phone calls to
mobile apps. The last extension on the monitor app was released on
July 8, 2022, and is an address search window that assists
newly-employed dispatchers who recently relocated to Atlanta.

\subsection{The Dashboard}
\label{subsect:dashboard}

\begin{figure}[htbp]
	\centering
	\includegraphics [width=0.75\linewidth] {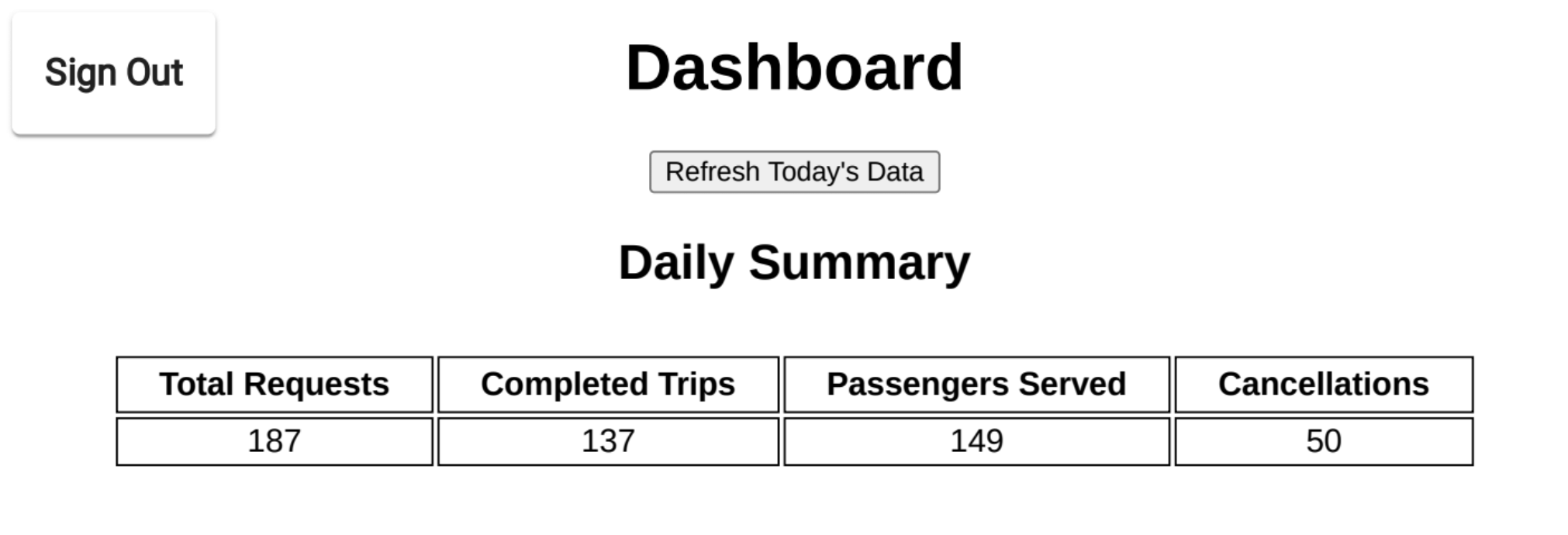}
	\caption{The Header of the Dashboard Web App.}
	\label{fig:dashboard}
\end{figure}

In addition to the live visualization provided by the monitor app, a
dashboard app was created to keep track of system performance over the
course of a day. The header of the dashboard is shown in Figure~\ref{fig:dashboard}.  The dashboard app displays trackers for total
requested trips, completed trips, and cancelations, both in aggregate,
and broken down by hour in bar graphs.  Tables with the details of
each trip, including the rider names, request times, board times,
drop-off times, vehicles taken, and zones of origin are displayed such
that one can identify riders who had issues, riders who made the same
trips each day, or riders who took multiple trips in a day. Riders who
canceled trips also have their cancellation time shown. All of these
features were introduced by May 4, 2022.  The dashboard also contains
a section tracking call-in requests for trips that were not made
through the app. Additionally, a small table tracked vehicles
that were removed from service due to driver inactivity. The last
table is a live updating section that provided the active status of
each online vehicle, whether it was idling, picking up a rider, or
serving a trip.  Beyond these metrics, a few map visualizations were
also included at the bottom of the dashboard.  The first map
visualization shows the origins and destinations of all trips over the
day within the shaded boundaries of the operating zones.  The second
map visualization displays the polylines of completed trips.  The
thickness of the polyline showing how frequently each road segment is
used.  A node.JS preservation script was also developed that would run
every day after service was over.  The preservation script would save
the day dashboard as a PDF and automatically upload that file to a
Dropbox which facilitated comparison of service quality over the
course of the pilot. The preservation script began service on May 25,
2022.

\subsection{Backend Servers and Infrastructure} \label{subsect:backend}

\begin{figure}[htbp]
	\centering
	\includegraphics [width=0.75\linewidth] {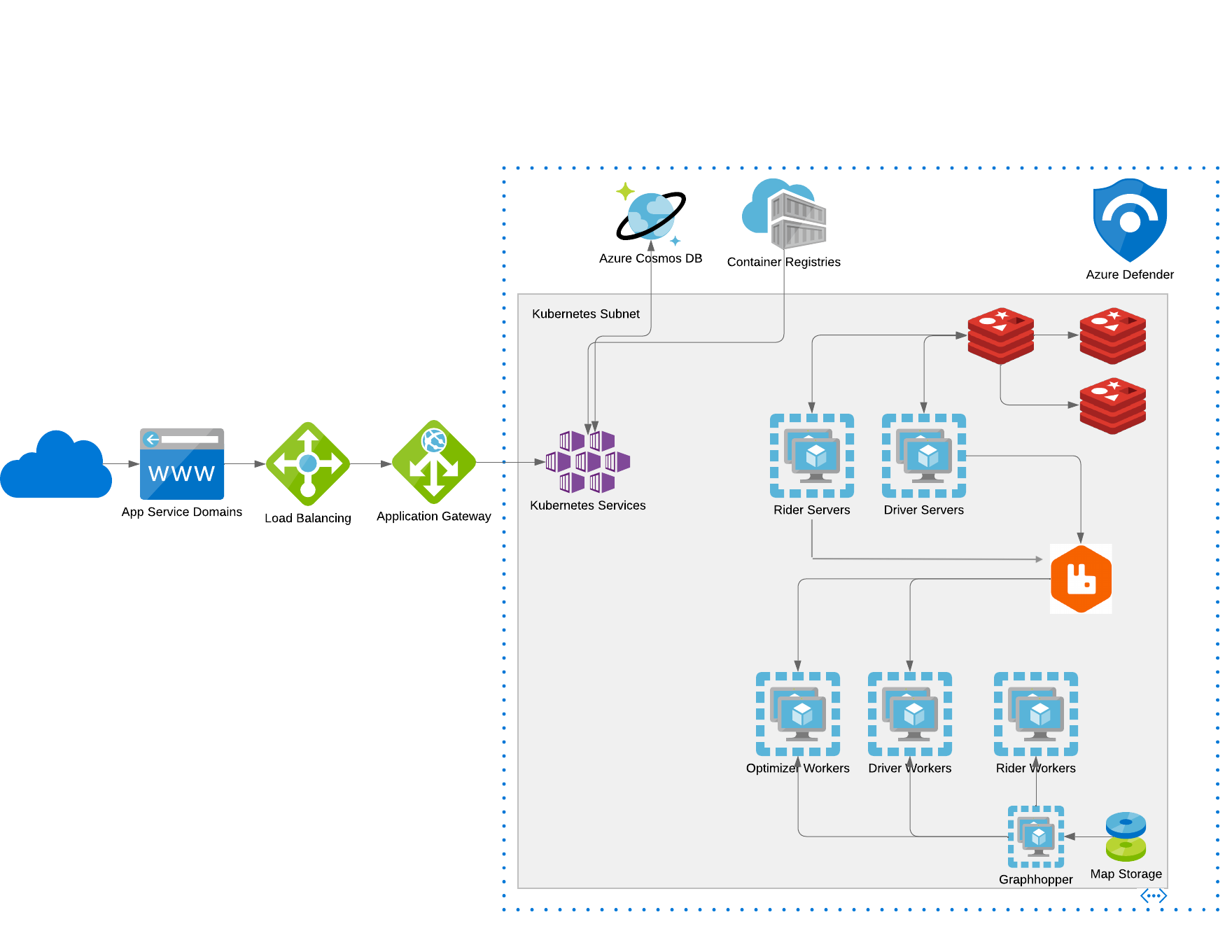}
	\caption{A Diagram of the Backend Infrastructure.}
	\label{fig:backend}
\end{figure}

The backend architecture consists of node.JS and Python servers, a
Redis cluster for pub/sub, the noSQL version of cosmosDB, and a
rabbitmq cluster.  A diagram of the backend is shown in
Figure~\ref{fig:backend}.  There are two node.JS servers: one for the
riders, and another for the drivers which manage authentication, user
sessions, and communication between the backend and the mobile apps.
Messages are passed to/from the backend from/to the mobile apps
through socket.io connections, which are subscribed to relevant Redis
queues.  For example, when a rider is assigned to a vehicle, the rider
app is subscribed to that vehicle's GPS-queue in Redis.  Thereafter,
whenever the driver app pushes a GPS coordinate to that queue, the
rider app will receive the new GPS position of the vehicle.  The
Python servers handle all the business logic, communicating with the
cosmosDB to make changes to the state of the system.  The Python
servers were split out by function, one handling rider functions,
another driver functions, and one per zone for functions that had to
be done sequentially per zone. Each Python server is managed by
Celery, a distributed task queue framework, which uses rabbitmq as a
message broker.  Each microservice was containerized and orchestrated
using Azure Kubernetes Service, which ensured multiple replications of
each server were running to provide maximal uptime.  Initially, the
backend control was limited to a state-of-the-art dispatching
algorithm, which was used to assign vehicles to passengers as ride
requests were made \citep{riley2022operating}.  On May 17, 2022 the
\textit{automatic vehicle removal} function to remove inattentive
drivers was added. On May 22, 2022, a rebalancing algorithm was deployed to reposition idle vehicles toward idle stops with higher predicted demand \citep{riley2022operating, lu2026impact}. It is worth noting that this rebalancing algorithm was developed iteratively during Phase~2 of the pilot, as the development team responded to operational observations in real time. The algorithm went through multiple incremental updates during the phase, and the version described in \citet{lu2026impact} represents its final state.

\section{Fleet Management and Driver Behavior}
\label{sec:fleet_and_driver}
\subsection{Shuttles}

\begin{figure}[htbp]
	\centering
	\includegraphics[width=0.65\textwidth]{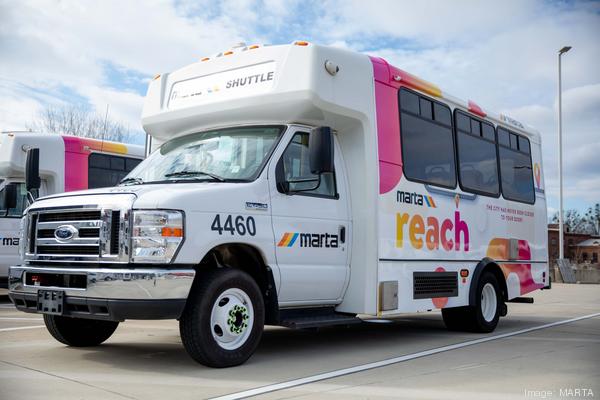}
	\caption{A \MartaReach{} Shuttle.}
	\label{fig:shuttle}
\end{figure}

\MartaReach{} relied on 18 branded \Reach{} vehicles, as pictured in Figure~\ref{fig:shuttle}.
The shuttles are easily recognizable to riders due to their branding and identifying numbers.
In fact, the branding was a critical aspect to make communities aware of the pilot, especially in reaching non-transit users who would not see the advertisements at MARTA bus and rail stations.
The shuttles were operated by drivers from two
contracting agencies (First Transit and Transdev) according to two
shifts: 6:00 a.m.\ to 1:00 p.m.\ and 1:00 p.m.\ to 7:00 p.m.

\subsection{Fleet Management}
\Reach{} used an operational protocol that required drivers to log into the system, receive requests, and follow instructions through the driver mobile application.
For safety reasons, drivers only interacted with the application when their vehicle was at a complete stop.
To ensure that the system operated properly, the fleet was monitored in real-time through the monitor web application.

During the pilot, {\em probably the most significant operational
	challenge was the delayed response to trip requests by drivers}.
Such delays led to long waiting times for riders and trip
cancellations. In Phase~1, dispatchers were authorized to manually
sign off unresponsive drivers through the monitor web application. These actions were
labeled as \emph{removed by admin}. In Phase~2,
the \Reach{} software was upgraded to enable automatic sign-off of
unresponsive vehicles by the server: if a driver with an empty vehicle
received an instruction but did not respond within a specific time, they were signed off the application and were no longer
considered for future requests until they signed on again. These events are marked \emph{removed by server} or
\emph{automatic vehicle removal} in the \Reach{} system. When
automatic vehicle removal was released on May 17, 2022, the time
threshold was set to five minutes, and it was reduced to four minutes on
June 4, 2022.
An exception was made for the last available vehicle in each zone to avoid having no active vehicles.

\subsection{Driver Behavior and Known Issues}
\label{subsect:driver_behavior}

\begin{figure}[htbp]
	\centering
	\includegraphics[trim={0 1.4cm 0 2cm}, clip, width=0.8\textwidth]{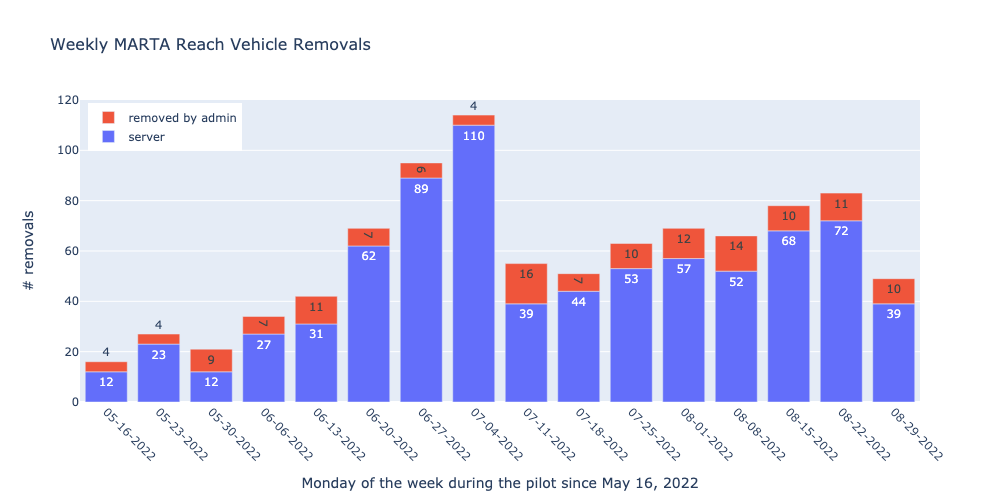}
	\caption{
		Weekly \MartaReach{} Vehicle Removals (tracked since May 16, 2022).
	}
	\label{fig:weekly_removal}
\end{figure}

As mentioned in Section~\ref{subsect:pilot_operation}, driver behavior and fleet management were two key challenges faced by \MartaReach{}.
This section provides data on these issues and describes (partial) solutions adopted in the pilot.

\subsubsection{Driver Response Time}
Recall that the \Reach{} platform has the ability to remove drivers
from the \Reach{} system, either manually through the monitor app or
automatically through the backend server. The frequency of vehicle
removals is reported in Figure~\ref{fig:weekly_removal} on a weekly
basis: they range from 16 to 114 per week, indicating that \Reach{}
drivers were frequently
unresponsive.
To quantify the response times, a particular scenario is considered where a vehicle was idle at a stop and needed to serve a single request without any intermediate stops.
This scenario
occurred 4,678 times during the pilot, and the mean and median
response times were 45 and 19 seconds, respectively. There were 396
and 65 instances where the driver reacted after more than two and five
minutes respectively.

The introduction of the Automatic Vehicle Removal function had
a significant impact on preventing long response times. Prior to its
implementation, the mean and median response times were 69 and 24
seconds respectively, with the longest response time reaching 1015
seconds (approximately 17 minutes). After the implementation of
Automatic Vehicle Removal, the mean and median response times were
reduced to 38 and 18 seconds respectively, with the longest waiting
time at 661 seconds. Note that the Automatic Vehicle Removal function
did not apply to the last remaining vehicle in a zone, which explains
this outlier.

\subsubsection{Shuttle Online Times}
\begin{figure}[htbp]
	\centering
	\includegraphics[width=\textwidth]{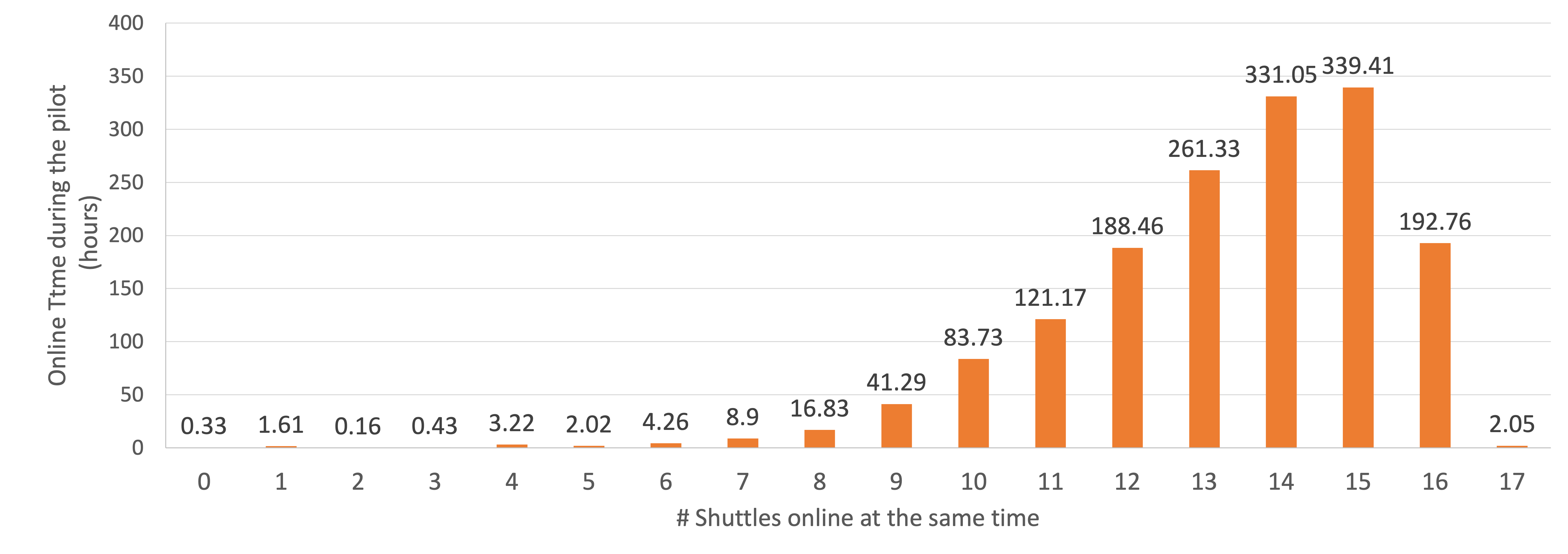}
	\caption{
		Fleet online hours during the pilot for different fleet-size
	}
	\label{fig:online_shuttle_count}
\end{figure}

The MARTA \Reach{} pilot had a fleet of 16 shuttles at its
disposal. The pilot spanned 132 service days with a daily working
schedule of 13 hours from 6:00 a.m.\ to 7:00 p.m.
The server experienced minor technical issues for nine
service dates from March 18, 2022 to March 30, 2022
which resulted in a few hours total of degraded service during that period.
Removing these days from the analysis, the drivers should have been signed in for
27,456 hours (123 fully recorded service days, 16 shuttles, and 13
hours each day per vehicle). However, in reality, the drivers were online for a total of
21,294 hours, i.e., 83.23\% of the planned vehicle online time.
This discrepancy can be
attributed to many reasons, including absences, driver shifts, breaks,
and driver shortages. Figure~\ref{fig:online_shuttle_count} provides
statistics on the number of working hours with different fleet sizes
during the pilot.  Under ideal conditions, the fleet should be at full strength for $123 \times 13 = 1,599$
service hours.
However, during the pilot, the \Reach{} fleet was fully active
for only 194.81 hours, or
12.82\% of the time.
For 29.52\% of the service hours no more than 75\% of the vehicles were online. By design, an ODMTS does not need every vehicle active at once to function well; instead, these metrics are meant to illustrate how difficult it was to use pilot resources effectively. These observations highlight the importance of human resource and fleet management, and suggest that better management practices and monitoring tools may help operate larger systems.
\begin{figure}[htbp]
	\centering
	\includegraphics[width=\textwidth]{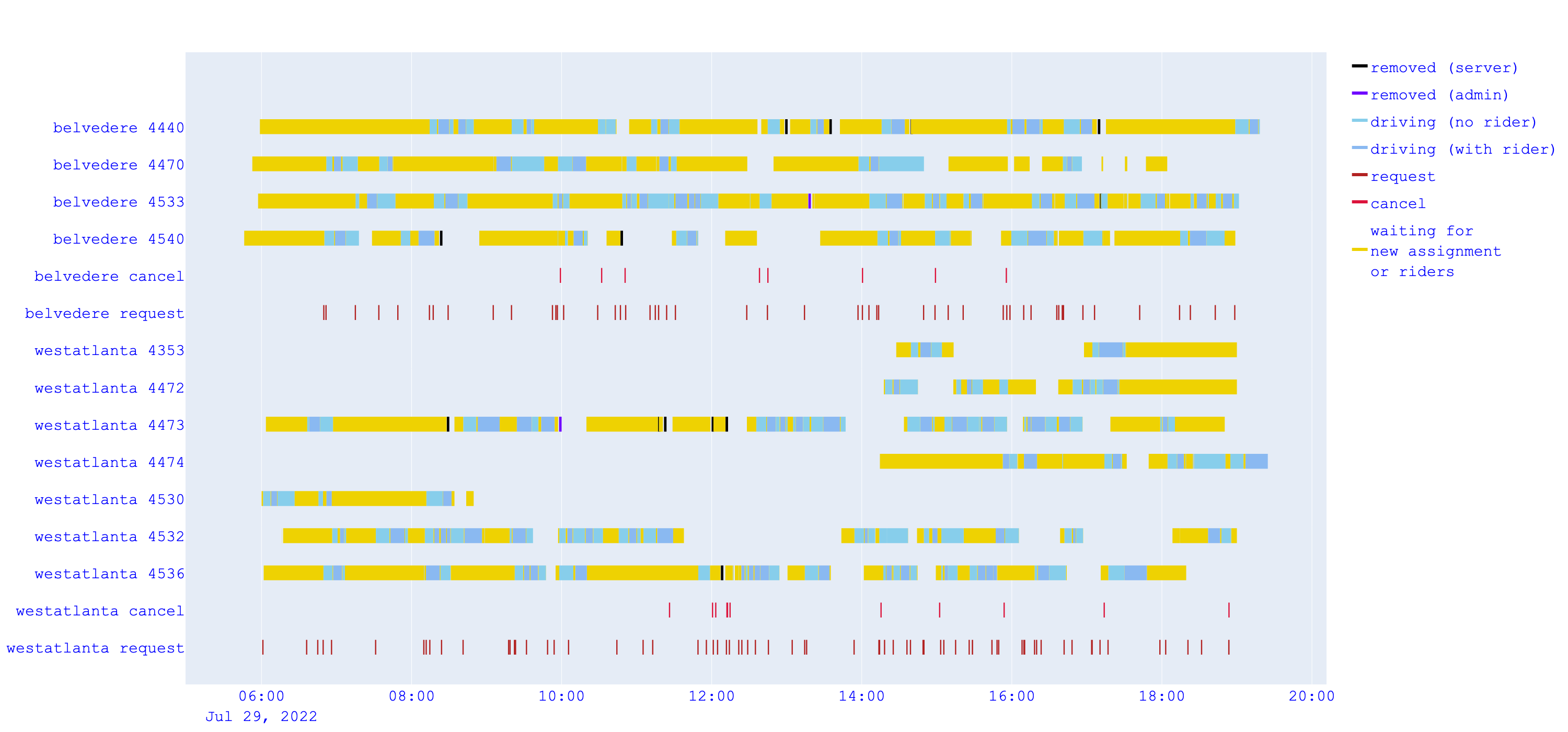}
	\caption{Gantt Chart of Vehicle Activities in West Atlanta and Belvedere on July 29, 2022.}
	\label{fig:one_day_fleet_acts_gantt_chart}
\end{figure}

Figure~\ref{fig:one_day_fleet_acts_gantt_chart} provides a Gantt chart
that shows the activities performed by \Reach{} vehicles in
Belvedere and West Atlanta on July 29, 2022. The chart highlights
major events, such as requests, cancellations, and vehicle removals,
and illustrates some of the driver and fleet management issues faced
during the pilot. Specifically, the Belvedere fleet usually had three
out of four vehicles online throughout the day. Around 2pm, all four
vehicles are serving passengers. Vehicles 4540 and 4470 were
particularly problematic that day with instances of removals followed
by long periods of inactivity (4540) and a long period of inactivity
at the end of the day (4470). The West Atlanta fleet started service
with four vehicles, which was then increased to six vehicles after 2pm
to address the higher travel demand. During most of the day, the West
Atlanta fleet was significantly short of the designated fleet-size
of six.
At noon, the fleet had only one or two active vehicles to serve a
considerable amount of requests. The rest of the vehicles were
signed out due to various reasons, indicating poor coordination
between drivers and dispatchers at shift times.
The absence of shuttles in the morning and around noon increased waiting times, despite the lower volume of requests at those times.
Addressing these fleet
management issues may further improve response times, trip durations,
and overall customer satisfaction. Shift management was particularly
lacking in the pilot, and is an important consideration for future
deployments.
While real-time monitoring can help flag these issues, part of the solution will likely be managerial.

\section{Brief of Optimization Models}
\label{sec:opt_models}

This sections summarizes the mathematical formulations of the two core algorithms underlying \MartaReach{}: the real-time dispatcher \citep{riley2019column} and the rebalancing algorithm \citep{lu2026impact}. Table~\ref{tab:notation_brief} defines the notation used throughout.

\begin{table}[th]
\centering
\caption{Summary of notation for the optimization models.}
\label{tab:notation_brief}
\footnotesize
\begin{tabular}{lp{0.65\textwidth}}
\toprule
\textbf{Notation} & \textbf{Description} \\
\midrule
\multicolumn{2}{l}{\textit{Sets}} \\[0.2em]
$S$ & Set of virtual stop locations (pickups and drop-offs). \\
$S_{\text{idle}} \subseteq S$ & Set of designated idle stops where shuttles may wait. \\
$R$ & Set of all candidate shuttle routes. \\
$V$ & Set of shuttles in the fleet. \\
$\hat{V}_{\text{idle}} \subseteq V$ & Set of idle shuttles not currently at an idle stop. \\
$N$ & Set of unserved requests carried over from previous epochs. \\
$N_L \subseteq N$ & Set of recent requests within the rolling consideration window. \\[0.4em]
\multicolumn{2}{l}{\textit{Parameters}} \\[0.2em]
$a_{rn}$ & 1 if route $r$ serves request $n$, 0 otherwise. \\
$c_r$ & Total waiting time accumulated by all passengers on route $r$. \\
$g_n$ & Time-dependent penalty for leaving request $n$ unserved; grows exponentially with waiting time. \\
$\gamma_s$ & Number of recent requests mapped to idle stop $s$. \\
$\rho_s^v$ & Travel time to reposition shuttle $v$ to idle stop $s$. \\[0.4em]
$p, d$ & Pickup and drop-off locations of a request. \\
$p_{idle}$ & Closest idle stop to $p$ in terms of travel time \\
\multicolumn{2}{l}{\textit{Decision variables}} \\[0.2em]
$y_r \in \{0,1\}$ & 1 if route $r$ is selected in the current epoch. \\
$x_n \in \{0,1\}$ & 1 if request $n$ remains unserved in the current epoch. \\
$\zeta_s \in \mathbb{N}^0$ & Number of shuttles to allocate to idle stop $s$. \\
$z_s$ & Proxy variable for $\zeta_s$ used in the rebalancing objective. \\
$w_s^v \in \{0,1\}$ & 1 if shuttle $v$ is repositioned to idle stop $s$. \\
\bottomrule
\end{tabular}
\end{table}

\subsection{Real-time dispatching}

The real-time dispatcher \citep{riley2019column} processes trip requests in discrete time epochs. At the end of each epoch, all pending unserved requests are batched and jointly optimized via a column generation algorithm, where each column represents a candidate route for one shuttle (an ordered sequence of pickup and drop-off stops a shuttle can follow to serve requests). This global batching approach avoids the locally-greedy decisions of nearest-vehicle assignment by simultaneously optimizing across all pending requests and all available shuttles, enabling ride-sharing and reducing system-wide wait times. The problem selects the best combination of routes:

\begin{align}
\min \quad & \sum_{r \in R} c_r y_r + \sum_{n \in N} g_n x_n \label{eq:rtdars_obj} \\
\text{s.t.} \quad & \sum_{r \in R} a_{rn}\, y_r + x_n = 1 \quad \forall n \in N \label{eq:rtdars_c1} \\
& \sum_{r \in R_v} y_r = 1 \quad \forall v \in V \label{eq:rtdars_c2} \\
& x_n,\, y_r \in \{0,1\} \quad \forall n \in N,\; r \in R \label{eq:rtdars_c3}
\end{align}

The objective \eqref{eq:rtdars_obj} minimizes total passenger waiting time across all selected routes, plus penalties for deferred requests; penalties grow exponentially to prioritize passengers who have waited longest. Constraint \eqref{eq:rtdars_c1} requires each request to be either served by some route or explicitly deferred to the next epoch. Constraint \eqref{eq:rtdars_c2} assigns every shuttle to exactly one route. Constraint \eqref{eq:rtdars_c3} enforces binary decisions on route selection and deferral. The model guarantees all requests are eventually served: deferred requests accumulate higher penalties until scheduled.

\subsection{Rebalancing}

When shuttles become idle, the rebalancing algorithm \citep{lu2026impact} repositions them to designated idle stops where demand is anticipated. The algorithm operates in two sequential stages.

\textbf{Stage 1: Fleet allocation.} For each idle stop $s$, recent demand is estimated by counting nearby requests in the rolling window $N_L$:

\begin{equation}
\gamma_s = \sum_{n \in N_L} \mathbb{I}\bigl[p^{\text{idle}}_n = s\bigr] \quad \forall s \in S_{\text{idle}}
\label{eq:gamma}
\end{equation}

The following model determines how many shuttles $\zeta_s$ to send to each stop:

\begin{align}
\min \quad & \sum_{s \in S_{\text{idle}}} \frac{\gamma_s}{z_s} \label{eq:mpc_obj} \\
\text{s.t.} \quad & z_s = \max\{0.5,\, \zeta_s\} \quad \forall s \in S_{\text{idle}} \label{eq:mpc_c1} \\
& \sum_{s \in S_{\text{idle}}} \zeta_s = |\hat{V}_{\text{idle}}| \label{eq:mpc_c2} \\
& \zeta_s \in \mathbb{N}^0 \quad \forall s \in S_{\text{idle}} \label{eq:mpc_c3}
\end{align}

The objective \eqref{eq:mpc_obj} minimizes the total demand-to-shuttle ratio across all idle stops, distributing the fleet proportionally to recent demand. Constraint \eqref{eq:mpc_c1} introduces the proxy $z_s$, which clips $\zeta_s$ at 0.5 to prevent division by zero and doubles the effective penalty for leaving a stop unoccupied. Constraint \eqref{eq:mpc_c2} ensures all idle shuttles not already at an idle stop are reassigned. Constraint \eqref{eq:mpc_c3} enforces non-negative integer allocations.

\textbf{Stage 2: Vehicle assignment.} Given the target counts $\zeta_s$ from Stage~1, the following MIP assigns specific shuttles to stops while minimizing total repositioning travel time:

\begin{align}
\min \quad & \sum_{v \in \hat{V}_{\text{idle}}} \sum_{s \in S_{\text{idle}}} \rho_s^v\, w_s^v \label{eq:mip_obj} \\
\text{s.t.} \quad & \sum_{v \in \hat{V}_{\text{idle}}} w_s^v = \zeta_s \quad \forall s \in S_{\text{idle}} \label{eq:mip_c1} \\
& \sum_{s \in S_{\text{idle}}} w_s^v = 1 \quad \forall v \in \hat{V}_{\text{idle}} \label{eq:mip_c2} \\
& w_s^v \in \{0,1\} \quad \forall s \in S_{\text{idle}},\; v \in \hat{V}_{\text{idle}} \label{eq:mip_c3}
\end{align}

The objective \eqref{eq:mip_obj} minimizes total deadhead rebalancing travel time across all relocating shuttles. Constraint \eqref{eq:mip_c1} ensures each stop receives exactly the number of shuttles specified by Stage~1. Constraint \eqref{eq:mip_c2} sends every idle shuttle to exactly one stop. Constraint \eqref{eq:mip_c3} enforces binary assignment decisions.

It is worth mentioning that the \Reach{} pilot divides this process into two separate stages rather than relying on a single unified optimization, for the following reasons. First, both models are relatively small in scale and can therefore be solved rapidly in real time. Second, formulating a single unified model would require effort to determine an appropriate weighting between the number of vehicles and vehicle travel time, which in turn would require historical data for proper calibration; such data was not available for the \Reach{} pilot. Finally, because each zone in the pilot is relatively small in size, even when a few shuttles are rebalanced and new requests subsequently arise, the fleet can still serve that demand promptly.

\end{document}